%% file: Xsll_arxiv_final_version.tex
\documentclass[twocolumn,showpacs,aps,prl,superscriptaddress]{revtex4}

\newcommand{\BABARPubYear}    {13}
\newcommand{\BABARPubNumber}  {01}

\newcommand{\SLACPubNumber} {15866}
\newcommand{\LANLNumber} {1312.5364}

\long\def\inst#1{\par\nobreak\kern 4pt\nobreak
 {\it #1}\par\vskip 10pt plus 3pt minus 3pt}

\usepackage[dvips]{graphicx}
\usepackage{epsf} 
\usepackage{epsfig} 
\usepackage{rotating}
\usepackage{color}
\usepackage{colordvi}
\usepackage{array}%
\usepackage{placeins}%

\newcommand{\psfile}[3][]{ 
 \begin{center}
 \setlength{\epsfxsize}{#3\linewidth}\leavevmode
 \def\noOpt{}\def\testit{#1}\ifx\testit\noOpt%
 \epsfbox{#2}%
 \else%
 \epsfbox[#1]{#2}%
 \fi
 \end{center}
}

\RequirePackage{xspace}

\usepackage{relsize}
\def\babar{\mbox{\slshape B\kern-0.1em{\smaller A}\kern-0.1em
 B\kern-0.1em{\smaller A\kern-0.2em R}}}

\def\epem {\ensuremath{e^+e^-}\xspace}

\def\mumu {\ensuremath{\mu^+\mu^-}\xspace}

\def\ellell {\ensuremath{\ell^+ \ell^-}\xspace}

\def\gaga {\ensuremath{\gamma\gamma}\xspace} 

\def\q {\ensuremath{q}\xspace}

\def\qqbar {\ensuremath{q\overline q}\xspace}

\def\ccbar {\ensuremath{c\overline c}\xspace}
\def\b {\ensuremath{b}\xspace}
\def\bbar {\ensuremath{\overline b}\xspace}
\def\bbbar {\ensuremath{b\overline b}\xspace}

\def\piz {\ensuremath{\pi^0}\xspace}

\def\pip {\ensuremath{\pi^+}\xspace}
\def\pim {\ensuremath{\pi^-}\xspace}
\def\pipi {\ensuremath{\pi^+\pi^-}\xspace}

\def\Kbar {\kern 0.2em\overline{\kern -0.2em K}{}\xspace}

\def\Kz {\ensuremath{K^0}\xspace}
\def\Kzb {\ensuremath{\Kbar^0}\xspace}
\def\KzKzb {\ensuremath{\Kz \kern -0.16em \Kzb}\xspace}
\def\Kp {\ensuremath{K^+}\xspace}
\def\Km {\ensuremath{K^-}\xspace}

\def\KpKm {\ensuremath{\Kp \kern -0.16em \Km}\xspace}
\def\KS {\ensuremath{K^0_{\scriptscriptstyle S}}\xspace}
\def\KL {\ensuremath{K^0_{\scriptscriptstyle L}}\xspace}

\def\Dbar {\kern 0.2em\overline{\kern -0.2em D}{}\xspace}

\def\Dz {\ensuremath{D^0}\xspace}
\def\Dzb {\ensuremath{\Dbar^0}\xspace}
\def\DzDzb {\ensuremath{\Dz {\kern -0.16em \Dzb}}\xspace}
\def\Dp {\ensuremath{D^+}\xspace}
\def\Dm {\ensuremath{D^-}\xspace}

\def\DpDm {\ensuremath{\Dp {\kern -0.16em \Dm}}\xspace}

\def\B {\ensuremath{B}\xspace}
\def\Bbar {\kern 0.18em\overline{\kern -0.18em B}{}\xspace}

\def\BB {\ensuremath{B\Bbar}\xspace}
\def\Bz {\ensuremath{B^0}\xspace}
\def\Bzb {\ensuremath{\Bbar^0}\xspace}
\def\BzBzb {\ensuremath{\Bz {\kern -0.16em \Bzb}}\xspace}
\def\Bu {\ensuremath{B^+}\xspace}
\def\Bub {\ensuremath{B^-}\xspace}

\def\BpBm {\ensuremath{\Bu {\kern -0.16em \Bub}}\xspace}

\def\BorBbar {\kern 0.18em\optbar{\kern -0.18em B}{}\xspace}
\def\DorDbar {\kern 0.18em\optbar{\kern -0.18em D}{}\xspace}
\def\KorKbar {\kern 0.18em\optbar{\kern -0.18em K}{}\xspace}

\def\jpsi {\ensuremath{{J\mskip -3mu/\mskip -2mu\psi\mskip 2mu}}\xspace}
\def\psitwos {\ensuremath{\psi{(2S)}}\xspace}

\mathchardef\Upsilon="7107
\def\Y#1S{\ensuremath{\Upsilon{(#1S)}}\xspace}

\def\FourS {\Y4S}

\mathchardef\Deltares="7101
\mathchardef\Xi="7104
\mathchardef\Lambda="7103
\mathchardef\Sigma="7106
\mathchardef\Omega="710A

\def\Deltabar{\kern 0.25em\overline{\kern -0.25em \Deltares}{}\xspace}
\def\Lbar{\kern 0.2em\overline{\kern -0.2em\Lambda\kern 0.05em}\kern-0.05em{}\xspace}
\def\Sigbar{\kern 0.2em\overline{\kern -0.2em \Sigma}{}\xspace}
\def\Xibar{\kern 0.2em\overline{\kern -0.2em \Xi}{}\xspace}
\def\Obar{\kern 0.2em\overline{\kern -0.2em \Omega}{}\xspace}
\def\Nbar{\kern 0.2em\overline{\kern -0.2em N}{}\xspace}
\def\Xb{\kern 0.2em\overline{\kern -0.2em X}{}\xspace}

\def\X {\ensuremath{X}\xspace}

\def\mes {\mbox{$m_{\rm ES}$}\xspace}

\def\DeltaE {\mbox{$\Delta E$}\xspace}

\newcommand{\tev}{\ensuremath{\mathrm{\,Te\kern -0.1em V}}\xspace}
\newcommand{\gev}{\ensuremath{\mathrm{\,Ge\kern -0.1em V}}\xspace}
\newcommand{\mev}{\ensuremath{\mathrm{\,Me\kern -0.1em V}}\xspace}
\newcommand{\kev}{\ensuremath{\mathrm{\,ke\kern -0.1em V}}\xspace}
\newcommand{\ev}{\ensuremath{\mathrm{\,e\kern -0.1em V}}\xspace}
\newcommand{\gevc}{\ensuremath{{\mathrm{\,Ge\kern -0.1em V\!/}c}}\xspace}
\newcommand{\mevc}{\ensuremath{{\mathrm{\,Me\kern -0.1em V\!/}c}}\xspace}
\newcommand{\gevcc}{\ensuremath{{\mathrm{\,Ge\kern -0.1em V\!/}c^2}}\xspace}
\newcommand{\mevcc}{\ensuremath{{\mathrm{\,Me\kern -0.1em V\!/}c^2}}\xspace}

\def\invfb {\ensuremath{\mbox{\,fb}^{-1}}\xspace}

\def\mus {\ensuremath{\rm \,\mus}\xspace}

\def\mus {\ensuremath{\,\mu{\rm s}}\xspace} 

\def\to  {\ensuremath{\rightarrow}\xspace}

\def\pep2{PEP-II}

\def\gsim{{~\raise.15em\hbox{$>$}\kern-.85em
 \lower.35em\hbox{$\sim$}~}\xspace}
\def\lsim{{~\raise.15em\hbox{$<$}\kern-.85em
 \lower.35em\hbox{$\sim$}~}\xspace}

\def\CP {\ensuremath{C\!P}\xspace}

\xspace

\def\jetset74 {\mbox{\tt Jetset \hspace{-0.5em}7.\hspace{-0.2em}4}\xspace}

\newcommand{\gevcccc}{\ensuremath{{\mathrm{\,Ge\kern -0.1em V^2\!/}c^4}}\xspace}

\def\Kmaybestar {\ensuremath{K^{(*)}\xspace}}

\def\kll {\B\to\Kmaybestar\ellell\xspace}

\def\kmaybell {\B\to\Kmaybestar\ellell\xspace}

\def\mll {\ensuremath{m_{\ell\ell}}\xspace}
\def\modeeight {\ensuremath{B^0\rightarrow \Kp \pim \mumu}\xspace}

\def\BToXll {\ensuremath{B\rightarrow X_{s}\, \ell^{+}\ell^{-}}\xspace}
\def\BToXee {\ensuremath{B\rightarrow X_{s}\, \epem}\xspace}
\def\BToXmm {\ensuremath{B\rightarrow X_{s}\, \mumu}\xspace}
\def\mll {\ensuremath{m_{\ell^+\ell^-}}}

\def\Xs {\ensuremath{X_s}\xspace}
\def\mx {\ensuremath{m_{X_s}}\xspace}
\def\mxone {\ensuremath{m_{X_s,1}}\xspace}
\def\mxtwo {\ensuremath{m_{X_s,2}}\xspace}
\def\mxthree {\ensuremath{m_{X_s,3}}\xspace}
\def\mxfour {\ensuremath{m_{X_s,4}}\xspace}

\def\lhr {\ensuremath{L_R}\xspace}
\def\nsig {\ensuremath{N_{\rm sig}}\xspace}

\def\nxfd {\ensuremath{N_{\rm xfd}}\xspace}
\def\nxfdsb {\ensuremath{N^{\rm SB}_{\rm xfd}}\xspace}
\def\nbb {\ensuremath{N_{B\Bbar}}\xspace}
\def\nbbsb {\ensuremath{N^{\rm SB}_{B\Bbar}}\xspace}
\def\ncc {\ensuremath{N_{\rm udsc}}\xspace}
\def\nccsb {\ensuremath{N^{\rm SB}_{\rm udsc}}\xspace}
\def\nchm {\ensuremath{N_{\rm chm}}\xspace}
\def\nchmsb {\ensuremath{N^{\rm SB}_{\rm chm}}\xspace}
\def\nhad {\ensuremath{N_{\rm had}}\xspace}
\def\nhadsb {\ensuremath{N^{\rm SB}_{\rm had}}\xspace}

\def\acp {\mbox{$A_{\CP}$}\xspace}
\def\acpccs {\mbox{$A_{\CP}^{\ccbar s}$}\xspace}

\def\prob {\ensuremath{\cal P}\xspace}

\usepackage{relsize}
\usepackage{xspace}

\begin{document}

\preprint{\babar-PUB-\BABARPubYear/\BABARPubNumber}
\preprint{SLAC-PUB-\SLACPubNumber}

\begin{flushleft}
SLAC-PUB-\SLACPubNumber\\
arXiv:\LANLNumber\\[10mm]
\end{flushleft}

\title{
 {\mathversion{bold}
 Measurement of the \BToXll branching fraction
 and search for direct CP violation
 from a sum of exclusive final states}
}

\input authors_sep2013.tex

\begin{abstract}
We measure the total branching fraction of the flavor-changing neutral-current process
\BToXll, along with partial branching fractions in bins of dilepton and hadronic system ($\Xs$) mass,
using a sample of $471 \times 10^{6}$~$\Upsilon(4S) \to \BB$ events
recorded with the \babar\ detector.
The admixture of charged and neutral $B$ mesons produced at \pep2\
are reconstructed by combining a dilepton pair
with 10 different $X_s$ final states.
Extrapolating from a sum over these exclusive modes, we measure a lepton-flavor-averaged inclusive branching fraction
${\cal B}(\BToXll)
  = \left(6.73^{+0.70}_{-0.64}[{\rm stat}]^{+0.34}_{-0.25}[{\rm exp~syst}] \pm 0.50[{\rm model~syst}]\right) \times 10^{-6}$
for $\mll^2 > 0.1 \gevcccc$.
Restricting our analysis exclusively to final states from which a decaying
$B$ meson's flavor can be inferred, we additionally report measurements of
the direct $\CP$ asymmetry $\acp$ in bins of dilepton mass;
over the full dilepton mass range, we find $\acp=0.04 \pm 0.11 \pm 0.01$
for a lepton-flavor-averaged sample.

\end{abstract}


\maketitle

\pagestyle{plain}

The $b \to s \ell^{+} \ell^{-}$ transition, where $b$
is a bottom quark, $s$ is a strange quark, and $\ellell$
is an $\epem$ or $\mumu$ pair, is forbidden at lowest
order in the standard model (SM) but is allowed at one
loop via electroweak penguin and $W$-box diagrams.
The amplitude for this decay is expressed
in terms of perturbatively calculable effective Wilson coefficients,
$C^{\rm eff}_{7}$,  $C^{\rm eff}_{9}$, and $C^{\rm eff}_{10}$,
which represent the electromagnetic penguin diagram,
and the vector part and the axial-vector part of the linear
combination of the $Z$ penguin and $W^+W^-$ box diagrams, respectively~\cite{Buchalla}.
Non-SM contributions can enter these loops at the same order
as the SM processes, modifying the Wilson coefficients from
their SM expectations and allowing experimental sensitivity to
possible non-SM physics~\cite{Altmannshofer:2011gn, Kosnik:2012dj, Drobnak:2011aa,DescotesGenon:2011yn, Oh:2011nb, Lee:2008xc, Ali:2002jg, Gambino:2004mv,acpnp:alok,acpnp:soni}.

We study the inclusive decay \BToXll, where $\X_s$ is a hadronic
system containing exactly one kaon,
using a sum over exclusive final states, which provides a basis for
extrapolation to the fully inclusive rate.
We measure the total branching fraction (BF), as well as partial
BFs in five disjoint dilepton mass-squared $q^{2} \equiv \mll^2$ bins and
four hadronic mass $\mx$ bins, which are defined in Table~\ref{tab:results}.
We additionally search for direct $\CP$ violation in the same $q^{2}$ bins.
The relative precision of our results is approximately a factor
of two better than the combined precision of all similar previously
published measurements~\cite{Beringer:1900zz}.

The $\X_s$ system in the lowest mass $\mx$ bin $\mxone$ contains a single
kaon with no other hadrons present; the $\mxtwo$ bin is populated only above the $K\pi$ threshold.
Results are also reported in an additional $q^{2}$ region $q^{2}_{0} \equiv 1 < q^{2} < 6 \gevcccc$,
i.e., the perturbative window away
from the photon pole at low $q^{2}$ and the $\ccbar$ resonances at higher $q^{2}$, where
theory uncertainties are well
controlled~\cite{Asatryan:2001zw, Asatryan:2002iy, Ghinculov:2002pe, Asatrian:2002va, Gambino:2003zm, Ghinculov:2003bx, Bobeth:2003at, Ghinculov:2003qd, Greub:2008cy, Huber:2007vv, Huber:2005ig, Beneke:2009az}.
The most recent SM predictions in this region are
${\cal B}^{\rm low}(\BToXmm) = (1.59 \pm 0.11) \times 10^{-6}$ and
${\cal B}^{\rm low}(\BToXee) = (1.64 \pm 0.11) \times 10^{-6}$~\cite{Huber:2007vv}.
Theory uncertainties in the $q^{2}$ range above the $\psitwos$ are also well-characterized but
relatively much larger than above, with SM predictions for $q^{2} > 14.4 \gevcccc$ of
${\cal B}^{\rm high}(\BToXmm) = (0.24 \pm 0.07) \times 10^{-6}$ and
${\cal B}^{\rm high}(\BToXee) = (0.21 \pm 0.07) \times 10^{-6}$~\cite{Huber:2007vv}.
The SM expectation in the $q^{2} > 4 m_{\mu}^{2}$ range is
${\cal B}(\BToXll) = (4.6 \pm 0.8) \times 10^{-6}$~\cite{Ghinculov:2003qd}.
Direct $\CP$ violation, defined as
$\acp \equiv (\mathrm{BF}_{\bbar} - \mathrm{BF}_{b}) / (\mathrm{BF}_{\bbar} + \mathrm{BF}_{b})$,
where $\b$ ($\bbar$) denotes a $\Bbar$ ($B$) parent,
is expected to be suppressed well below the $1\%$ level in both exclusive and inclusive $b \to s \ellell$
transitions~\cite{acpsm:du,acpsm:ali,Bobeth:2008ij,Altmannshofer:2008dz};
however, in beyond-SM models with four quark generations, significant enhancements
are possible, particularly in the high-$q^{2}$ region~\cite{acpnp:alok,acpnp:soni}.

The \babar\,~\cite{Aubert:2004it} and Belle~\cite{Iwasaki:2005sy}
Collaborations have previously published \BToXll BFs
based on a sum over exclusive final states using only $\sim25\%$ of each
experiment's final dataset.
More recently, both collaborations (along with LHCb and CDF) have
published BFs, and time-integrated rate and angular asymmetries, for the exclusive decays
$\kll$~\cite{Lees:2012tva, Wei:2009zv, Aaltonen:2011qs, Aaij:2011aa, Aaij:2013qta, Aubert:2008bi, acp:lhcb}.
The present analysis uses the $424.2 \pm 1.8 \invfb$ $\epem \to \FourS$
data sample~\cite{Lees:2013rw}, corresponding to $\sim 471$
million $\BB$ pairs, collected with the \babar\
detector~\cite{BabarDetector,BabarDetectorUpdate}
at the \pep2\ collider at the SLAC National Accelerator Laboratory.

The decays \BToXll\ are reconstructed in 10 separate $\Xs$ hadronic final states
($K^+$, $K^+ \pi^0$, $K^+ \pi^-$, $K^+ \pi^- \pi^0$, $K^+ \pi^- \pi^+$,
$\KS$, $\KS \pi^0$, $\KS \pi^+$, $\KS \pi^+ \pi^0$, and $\KS \pi^+ \pi^-$)~\cite{CC},
combining these with an $\epem$ or $\mumu$ pair for a total of 20 final states.
The selection of charged and neutral particle candidates, as
well as the reconstruction of $\piz \to \gaga$ and $\KS \to \pipi$,
is described in Refs.~\cite{Aubert:2008bi, Lees:2012tva}.
Based on studies including up to 18 $\Xs$ modes with a maximum of
four pions and $\mx$ as large as $2.2\gevcc$,
we limit the number of $\Xs$ final states to the 10 listed above and require $\mx<1.8 \gevcc$
since the expected signal-to-background ratio rapidly decreases
with increasing $\Xs$ pion multiplicity and mass.
We assume that the fraction of modes containing a \KL\ is equal to that containing
a \KS and account for these decays, as well as $\KS \to \piz \piz$ and $\piz$ Dalitz
decays, in our reconstruction efficiencies.
With these efficiencies taken into account, the reconstructed
states represent $\sim70\%$ of the total inclusive rate.

We account for missing hadronic final states,
as well as for states with $\mx > 1.8 \gevcc$,
based on the formalism of Refs.~\cite{Kruger:1996cv,Ali:1996bm,Ali:2002jg,Asatryan:2001zw,Bobeth:1999mk,Huber:2007vv},
with hadronization of the $\Xs$ system provided by the
JETSET~\cite{Sjostrand:1993yb} event generator.
Given that we observe no statistically significant non-resonant $B \to K\pi \ellell$
decays in our data~\cite{Lees:2012tva},
signal decays with a two-body $\Xs$ system and $\mx < 1.1 \gevcc$
are assumed to proceed through the $K^{*}(892)$ resonance.
The simulation of such events, as well as those with a single kaon
and no pions, is similar to that for inclusive events but
incorporates the form factor models of Refs.~\cite{Ball:2004rg,Ball:2004ye}.

The kinematic variables
$\mes=\sqrt{E^2_{\rm CM}/4 -p^{*2}_B}$ and
$\DeltaE = E_B^* - E_{\rm CM}/2$,
where $p^*_B$ and $E_B^*$ are the $B$
momentum and energy in the $\FourS$
center-of-mass (CM) frame with
$E_{\rm CM}$ the total CM energy,
are used to distinguish signal from background events.
We require $\mes > 5.225 \gevcc$ and
$-0.10 < \DeltaE < 0.05 \gev$ ($-0.05 < \DeltaE < 0.05 \gev$)
for dielectron (dimuon) final states.
Signal-like $B$ backgrounds with $\jpsi$ ($\psitwos$) daughters
are removed by vetoing events with $6.8 < q^{2} < 10.1 \gevcccc$ ($12.9 < q^{2} < 14.2$).
We reconstruct $\Xs h^{\pm} \mu^{\mp}$ final states,
where $h$ is a track with no particle identification (PID) requirement applied,
to characterize backgrounds from hadrons misidentified as muons.
Such backgrounds occur only in dimuon
final states because of the significantly higher probability
to misidentify $\Kp$ or $\pip$ as a muon rather than an electron.
Similarly, backgrounds from $\B \to D (\to K^{(*)} \pi) \pi$ decays
occur only in dimuon modes and, assigning the pion mass hypothesis to both muon candidates,
we reject candidates with $K^{(*)} \pi$ mass values in the range $1.84 < m_{K^{(*)} \pi} < 2.04
\gevcc$.

We suppress $\epem \to \qqbar$ events (where $q$ is a
$u$, $d$, $s$ or $c$ quark) and $\BB$ combinatoric backgrounds
using boosted decision trees (BDTs)~\cite{Breiman,Narsky}
identical in construction to those used in our $\kll$
analysis~\cite{Lees:2012tva}. These BDTs are respectively
trained with simulated $udsc$ or $\BB$ backgrounds
and correctly reconstructed signal events.
Ensembles of simulated event samples are used to simultaneously optimize
the $\DeltaE$ windows and selection on the $udsc$ BDTs
for each individual $q^{2}$ and $\mx$ bin.
After all selection criteria are applied,
the average multiplicity of $B$ candidates
per event is $\approx$2.6 ($\approx$2.2)
for $\epem$ $(\mumu)$ final states.
We allow only one candidate per event,
selecting the candidate with the smallest
$|\DeltaE|$.
Signal efficiencies after
event selection
range from about $1$ to $30\%$ depending on mode
and the $q^{2}$ or $\mx$ bin.

In each $q^{2}$ and $\mx$ bin, we extract the signal yield
with a two-dimensional maximum likelihood (ML) fit
using $\mes$ and a likelihood ratio $\lhr$
based on the $\BB$ BDT,
$\lhr \equiv \prob_{\rm S} / (\prob_{\rm S} + \prob_{\rm B})$,
where $\prob_{\rm S}$ and $\prob_{\rm B}$ are, respectively,
probabilities for genuine-signal and $\BB$ backgrounds.
For correctly reconstructed signal events, $\lhr$ sharply peaks
near one, while $\BB$ backgrounds peak at zero. Events
with $\lhr > 0.42$ are selected.
This selection rejects $\gtrsim 95\%$
of the $\BB$ background events remaining after all other event
selections have been applied, with only a trivial reduction in signal
efficiency.

Five (six) event classes contribute to the dielectron (dimuon) ML fit:
(1) correctly reconstructed signal;
(2) events that contain a partially or incorrectly reconstructed
\BToXll decay (signal cross-feed);
(3) $udsc$ and (4) $\BB$ combinatorial backgrounds;
(5) charmonium backgrounds;
and, for dimuon modes, (6) events with hadrons
misidentified as muons.

There is no correlation between $\mes$ and $\lhr$ for
correctly reconstructed signal events. Therefore, the
probability distribution function (PDF) for these events
is chosen as a product of two one-dimensional (1D) PDFs, with $\mes$
parameterized with a Crystal Ball (CB)
function~\cite{Oreglia:1980cs,Gaiser:1982yw,Skwarnicki:1986xj}
and $\lhr$ described by a non-parametric histogram PDF.
The CB shape parameters are fixed using simulated signal events,
as is the $\lhr$ PDF.
These PDFs describe well the
$\mes$ and $\lhr$ distributions derived from the
high-statistics control samples of vetoed signal-like charmonium events.
The signal cross-feed is modeled as a two-dimensional (2D) $\mes$ versus $\lhr$ histogram PDF
using simulated signal samples, with normalization $\nxfd$
scaled as a fixed fraction of the fit signal yield $\nsig$.

The $udsc$ combinatoric background PDF is derived from simulated events
using a 2D non-parametric kernel density estimator with
adaptive bandwidth~\cite{Parzen,Epanechnikov,Narsky}, which
is validated using data collected with $\epem$ center-of-mass energy
$40\mev$ below the $\FourS$ resonance.
The $udsc$ normalization $\ncc$ is obtained by scaling the $43.9 \pm 0.2 \invfb$ of
off-resonance data~\cite{Lees:2013rw} by the ratio of on- to off-resonance integrated
luminosity.

The shape of the 2D PDF for the $\BB$ combinatoric background is modeled similarly to
the $udsc$ background. Its normalization in the
$5.225 < \mes < 5.270 \gevcc$ sideband, where no correctly reconstructed
signal events are expected, is obtained by subtracting
the $\nxfdsb$, $\nccsb$, $\nchmsb$ and $\nhadsb$ (for dimuon events)
contributions from the total number of sideband events,
giving the $\BB$ yield in the sideband region $\nbbsb$.
We use simulated events to obtain the ratio of the number of
$\BB$ combinatoric events in the $\mes>5.27 \gevcc$ signal
region to the number in the sideband region to scale $\nbbsb$
into the expected contribution $\nbb$ in the full fit region.

Charmonium backgrounds escaping the vetoed $q^{2}$ regions
are similarly described by a 2D kernel estimator, with
normalization $\nchm$ derived from a fit to the data in the vetoed
regions that is extrapolated into the non-vetoed regions.
The normalization $\nhad$ and shape of the 2D PDF for misidentified dimuon events
are characterized by a weighted 2D histogram taken directly from data
using event-by-event weights obtained from PID control samples~\cite{Aubert:2006vb,Lees:2012tva}.

\begin{table*}
\caption{\BToXee, \BToXmm and \BToXll partial BFs (in units
of $10^{-6}$) and $\acp$ by $q^{2} (\gevcccc)$ and $\mx (\gevcc)$ bin.
The number in parentheses after each result is the multiplier which is applied to
the measured semi-inclusive rate to account for unreconstructed and $\mx > 1.8 \gevcc$ final states.
Estimated contributions from the vetoed charmonium $q^{2}$ regions are included in both the total and $\mx$ binned results,
but not in the total $\acp$.
The first uncertainties are statistical, the second experimental systematics
and the third model-dependent systematics associated with the multiplicative factor.
There are no model-dependent $\acp$ systematics and $\acp$ is not measured as a function of $\mx$; the multiplicative factors are not used in calculating the total $\acp$.}
\centering\footnotesize
\setlength{\tabcolsep}{5pt}
\addtolength{\extrarowheight}{2pt}
\begin{tabular}{cccccc}
\hline \hline
Bin         & Range             & \BToXee                                                  & \BToXmm                                                  & \BToXll                                               & $\acp_{\BToXll}$    \\
\noalign{\vskip 0.5mm}
\hline
\noalign{\vskip 1mm}
$q^{2}_{0}$ & $1.0 < q^{2} < 6.0$   & $1.93^{+0.47}_{-0.45}{}^{+0.21}_{-0.16} \pm 0.18$ (1.71) & $0.66^{+0.82}_{-0.76}{}^{+0.30}_{-0.24} \pm 0.07$ (1.78) & $1.60^{+0.41}_{-0.39}{}^{+0.17}_{-0.13} \pm 0.18$ & $-0.06 \pm 0.22 \pm 0.01$ \\
$q^{2}_{1}$ & $0.1 < q^{2} < 2.0$   & $3.05^{+0.52}_{-0.49}{}^{+0.29}_{-0.21} \pm 0.35$ (1.96) & $1.83^{+0.90}_{-0.80}{}^{+0.30}_{-0.24} \pm 0.20$ (2.02) & $2.70^{+0.45}_{-0.42}{}^{+0.21}_{-0.16} \pm 0.35$ & $-0.13 \pm 0.18 \pm 0.01$ \\
$q^{2}_{2}$ & $2.0 < q^{2} < 4.3$   & $0.69^{+0.31}_{-0.28}{}^{+0.11}_{-0.07} \pm 0.07$ (1.73) &$-0.15^{+0.50}_{-0.43}{}^{+0.26}_{-0.14} \pm 0.01$ (1.80) & $0.46^{+0.26}_{-0.23}{}^{+0.10}_{-0.06} \pm 0.07$ & $\: 0.42 \: \: {}_{-0.42}^{+0.50} \pm 0.01$ \\
$q^{2}_{3}$ & $4.3 < q^{2} < 6.8$   & $0.69^{+0.31}_{-0.29}{}^{+0.13}_{-0.10} \pm 0.05$ (1.53) & $0.34^{+0.54}_{-0.50}{}^{+0.19}_{-0.15} \pm 0.03$ (1.59) & $0.60^{+0.27}_{-0.25}{}^{+0.10}_{-0.08} \pm 0.05$ & $\! \! -0.45_{-0.57}^{+0.44} \pm 0.01$ \\
$q^{2}_{4}$ & $10.1< q^{2} <12.9$   & $1.14^{+0.42}_{-0.40}{}^{+0.22}_{-0.10} \pm 0.04$ (1.16) & $0.87^{+0.51}_{-0.47}{}^{+0.11}_{-0.08} \pm 0.03$ (1.18) & $1.02^{+0.32}_{-0.30}{}^{+0.10}_{-0.07} \pm 0.04$ & \\
$q^{2}_{5}$ & $14.2< q^{2}$         & $0.56^{+0.19}_{-0.18}{}^{+0.03}_{-0.03} \pm 0.00$ (1.02) & $0.60^{+0.31}_{-0.29}{}^{+0.05}_{-0.04} \pm 0.00$ (1.02) & $0.57^{+0.16}_{-0.15}{}^{+0.03}_{-0.02} \pm 0.00$ & \\
$q^{2}_{45}$&$q^{2}_{4} \cup q^{2}_{5}$& ---                                                      & ---                                                      & ---                                               & $0.19 \: \: {}_{-0.17}^{+0.18} \pm 0.01$ \\
\noalign{\vskip 1mm}
\hline
\noalign{\vskip 1mm}
$\mxone$    & $0.4 < \mx < 0.6$ & $0.69^{+0.18}_{-0.17}{}^{+0.04}_{-0.03} \pm 0.00$ (1.00) & $0.74^{+0.25}_{-0.23}{}^{+0.04}_{-0.04} \pm 0.00$ (1.00) & $0.71^{+0.15}_{-0.14}{}^{+0.03}_{-0.03} \pm 0.00$ & \\
$\mxtwo$    & $0.6 < \mx < 1.0$ & $1.20^{+0.34}_{-0.33}{}^{+0.10}_{-0.07} \pm 0.00$ (1.00) & $0.76^{+0.44}_{-0.40}{}^{+0.08}_{-0.07} \pm 0.00$ (1.00) & $1.02^{+0.27}_{-0.25}{}^{+0.06}_{-0.05} \pm 0.00$ & \\
$\mxthree$  & $1.0 < \mx < 1.4$ & $1.60^{+0.72}_{-0.69}{}^{+0.27}_{-0.19} \pm 0.05$ (1.18) & $0.65^{+1.16}_{-1.08}{}^{+0.27}_{-0.25} \pm 0.02$ (1.18) & $1.32^{+0.61}_{-0.58}{}^{+0.19}_{-0.15} \pm 0.05$ & \\
$\mxfour$   & $1.4 < \mx < 1.8$ & $1.88^{+0.76}_{-0.73}{}^{+0.71}_{-0.47} \pm 0.12$ (1.91) & $0.19^{+1.35}_{-1.25}{}^{+0.70}_{-0.50} \pm 0.10$ (1.91) & $1.36^{+0.67}_{-0.63}{}^{+0.50}_{-0.34} \pm 0.12$ & \\
\noalign{\vskip 1mm}
\hline \hline
\noalign{\vskip 1mm}
Total       & $0.1 < q^{2}$         & $7.69^{+0.82}_{-0.77}{}^{+0.50}_{-0.33} \pm 0.50$        & $4.41^{+1.31}_{-1.17}{}^{+0.57}_{-0.42} \pm 0.27$        & $6.73^{+0.70}_{-0.64}{}^{+0.34}_{-0.25} \pm 0.50$ & $0.04 \pm 0.11 \pm 0.01$ \\
\noalign{\vskip 1mm}
\hline \hline
\end{tabular}
\label{tab:results}
\end{table*}

We extract the $\nsig$ central value and associated upper and lower
limits using the negative log-likelihood (NLL) for $\nsig$. We calculate
partial BFs taking into account the efficiency for each
final state in each $q^{2}$ and $\mx$ bin, as well as the multiplicative factors
that provide extrapolation to the fully inclusive BFs. The results
are shown in Table~\ref{tab:results}, where the fully inclusive total rate and the
$\mx$ binned results include estimated signal contributions in the vetoed charmonium $q^{2}$ regions.
Fit projections for all $q^{2}$ and $\mx$ bins
are available as supplemental EPAPS material~\cite{EPAPS}, along with
a table giving the raw numerical results from our fits.
Figure~\ref{fig:results} shows our $q^2$ binned results overlaid on the nominal SM
expectations derived from our $\BToXll$ signal model. A similar plot for $\mx$
is included as supplemental material.

We consider systematic uncertainties associated with
purely experimental systematic uncertainties and the model-dependent
extrapolation to the fully inclusive rate.
The experimental systematics can either be additive,
affecting the extraction of the signal yield from the data,
or multiplicative, affecting the calculation
of a BF from an observed signal yield.
Sources of multiplicative systematic uncertainty include $\BB$ counting as well as
tracking, PID and reconstruction efficiencies. The only significant additive systematic
uncertainties are associated with the PDF parameterizations and normalizations.
The total experimental systematic uncertainty is the sum-in-quadrature
of the above terms, with the exception that uncertainties related to charged particle tracking efficiencies are
assumed to be fully correlated among all charged particles.
The evaluation of all experimental systematics is fully described in Ref.~\cite{Lees:2012tva}.
Tables quantifying each individual contribution to the experimental and model-dependent
extrapolation systematic uncertainties are available as supplemental EPAPS material~\cite{EPAPS}.

The uncertainty in the extrapolation to the inclusive rate is characterized
through variations that attempt to quantify
our lack of knowledge of the true dilepton mass-squared distribution and hadronization of the $X_s$ system
beyond the specific final states and $\mx$ range that we observe.
We average the most recent $\kll$ BFs~\cite{Amhis:2012bh},
excluding $\babar$ results, and use the latest \babar\ result~\cite{Aubert:2004ur}
for the ratio of charged-to-neutral $\FourS \to \bbbar$ decays,
$\Gamma(\BpBm) / \Gamma(\BzBzb) = 1.006 \pm 0.036 \pm 0.031$.
Each of these terms is varied by its one-standard-deviation uncertainty.
We examine an alternate $\mx$ transition point of $1.0\gevcc$
between the $\kll$ and $\BToXll$ models.
To account for hadronization uncertainties in $\mx > 1.1 \gevcc$ events,
we generate 20 simulated datasets with varied JETSET tunings, two
different values for the $B$-meson Fermi motion, and two different
$b$-quark mass values. We take the full spread of the extrapolation factors
derived from these variations to estimate this systematic uncertainty.
Additionally, for $\mx > 1.1 \gevcc$,
the fraction of modes with more than
one $\piz$ is varied around the generator
value of 0.20 by $\pm 50\%$;
the fraction of modes with either no $\piz$ and more
than two charged pions, or one $\piz$ and more than
one charged pion, is varied by $\pm 50\%$ around the
$q^{2}$-dependent generator value;
and the fraction of modes with more than one
neutral or charged kaon is varied around the
generator value of 0.034 by $\pm 50\%$.
Contributions from final states with photons
that do not come from $\piz$ decays but rather
from $\eta$, $\eta^\prime$, $\omega$, etc., are
expected to be insignificant, and we do not
vary the fractions of these decays. Each
of the above variations is added in quadrature
to obtain the final model-dependent systematic.
Table~\ref{tab:results} lists both the
experimental and model-dependent systematics.

\begin{figure}
  \centering
  \includegraphics[width=\columnwidth]{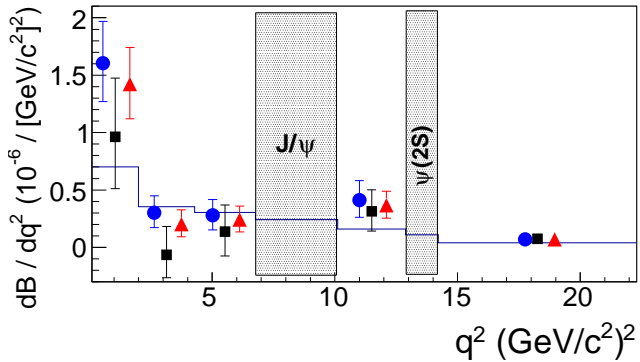}
  \caption{
    Differential BF as a function of $q^{2}$ for
    electron (blue circles), muon (black squares) and
    lepton-flavor-averaged final states (red triangles). The
    errors correspond to the total uncertainties.
    The histogram shows the SM expectation, which has
    uncertainties of approximately 10-30\% in different $q^{2}$ regions.
    The shaded boxes denote the vetoed charmonium regions. The horizontal
    spread of points in each bin is meant only to aid visibility.
  }\label{fig:results}
\end{figure}

We calculate the total inclusive rate by summing the $q^{2}_{1}$
through $q^{2}_{5}$ results taking into account correlations
in the systematic uncertainties and estimating signal contributions
in the vetoed charmonium $q^{2}$ regions.
The lepton-flavor-averaged $\BToXll$ results are weighted averages of
the individual $\BToXee$ and $\BToXmm$ results that take into account
correlations in the systematic uncertainties.
Figure~\ref{fig:results} shows the
differential BF results as a function of $q^{2}$ and $\mx$ overlaid
with the SM expectation.
The results in these bins,
as well as in the $q^{2}_{0}$ region,
are generally in good agreement with SM predictions.
Given our experimental uncertainties, we are insensitive to the relatively
small differences in the $\epem$ and $\mumu$ rates expected in the SM,
and observe no significant differences between $\epem$ and $\mumu$ final states.

Several model-independent analyses of the
form-factor-independent angular observables reported
in a recent $\modeeight$ LHCb analysis~\cite{Aaij:2013qta}
explain the anomaly reported there in terms of a non-vanishing
beyond-SM contribution $C^{\rm BSM}_{9}$~\cite{Descotes-Genon:2013wba,Jager:2012uw,Buras:2013qja,Altmannshofer:2013foa,Gauld:2013qba,Gauld:2013qja,Beaujean:2013soa,Horgan:2013pva,Buras:2013dea,Hurth:2013ssa}.
These phenomenological studies all present generally similar results,
yielding a three-sigma range for $C^{\rm BSM}_{9}$ of $\sim [-2,0]$,
implying a corresponding suppression in the fully inclusive
BF of up to $\sim 25\%$ in the $1 < q^{2} < 6 \gevcccc$
and $q^{2} > 14.4 \gevcccc$ ranges.
Although our results in the $q^{2}_{0}$ range are consistent with both the SM expectation
as well as a possible suppression in the decay rate,
our results in the $q^{2}_{5}$ range show an excess, rather than a deficit, of $\sim 2 \sigma$
in both the $\BToXee$ and $\BToXmm$ rates with respect to the SM expectation~\cite{Huber:2007vv}.

We search for $\CP$ violation in each $q^2$ bin by dividing our
dataset into four disjoint samples according to lepton identity ($\epem$ or $\mumu$) and
the $B$ or $\Bbar$ flavor as
determined by the kaon and pion
charges of the $\Xs$ system. Modes with $\Xs=\KS$,
$\KS\piz$ or $\KS\pipi$ are not used; and, because we perform no
model-dependent extrapolation of signal rates, we measure $\acp$ only
for the particular combination of final states used here.
We simultaneously fit all four datasets,
sharing a single value of $\acp$ as a free parameter,
using the BFs fit model described above.
Our $\acp$ results are shown in Table~\ref{tab:results};
a plot of the results as a function of $\q^2$ is included as part of
our supplemental EPAPS material~\cite{EPAPS}.
We analyze the vetoed
$\jpsi$ dataset, where $\CP$ violation is
expected to be trivially small~\cite{Wu:1999nc,Hou:2006du}, with the same fitting
methodology used for the signal $q^2$ bins;
we find $\acpccs = 0.0046 \pm 0.0057 [{\rm stat}]$. Observing no significant bias, we assign
the statistical uncertainty here as the systematic uncertainty for the $\acp$ results.
To extract $\acp$ for the full dilepton mass range, we sum the $\acp$
BFs across the four disjoint $\acp$ $q^2$ bins;
excluding the charmonium veto windows, we find
$\acp=0.04 \pm 0.11[{\rm stat}] \pm 0.01[{\rm syst}]$.
We observe no significant asymmetry in any
$q^2$ region or for the full dilepton mass range.

In summary, we have measured the total and partial BFs, as well as $\acp$,
for the inclusive radiative electroweak process $\BToXll$.
Our results are in general agreement with SM expectations
with the exception of our partial BF results in the high-$q^2$ region,
which show a $\sim 2 \sigma$ excess compared to both the SM expectation
and the most favored value of the beyond-SM contribution $C^{\rm BSM}_{9}$
advanced to explain recent observations by LHCb~\cite{Aaij:2013qta}.

We are grateful to Enrico Lunghi, Tobias Hurth and Tobias Huber
for useful discussions, as well as providing dilepton
mass-squared theory distributions derived using the most
up-to-date corrections.
We are additionally grateful for the excellent luminosity and machine conditions
provided by our \pep2\ colleagues,
and for the substantial dedicated effort from
the computing organizations that support \babar.
The collaborating institutions wish to thank
SLAC for its support and kind hospitality.
This work is supported by
DOE and NSF (USA), NSERC (Canada), CEA and CNRS-IN2P3 (France),
BMBF and DFG (Germany), INFN (Italy), FOM (The Netherlands),
NFR (Norway), MES (Russia), MINECO (Spain), STFC (United Kingdom).
Individuals have received support from the Marie Curie EIF
(European Union) and the A.~P.~Sloan Foundation (USA).


\clearpage

\title{
 {\mathversion{bold}
 Supplemental Material for ``Measurement of the \BToXll branching fraction
 and search for direct $\CP$ violation from a sum of exclusive final states''}
}

\maketitle

\pagestyle{plain}

\onecolumngrid

\appendix

\section{Introduction}
\label{app:intro}

\begin{figure}[!h]
  \centering
  \includegraphics[width=\columnwidth]{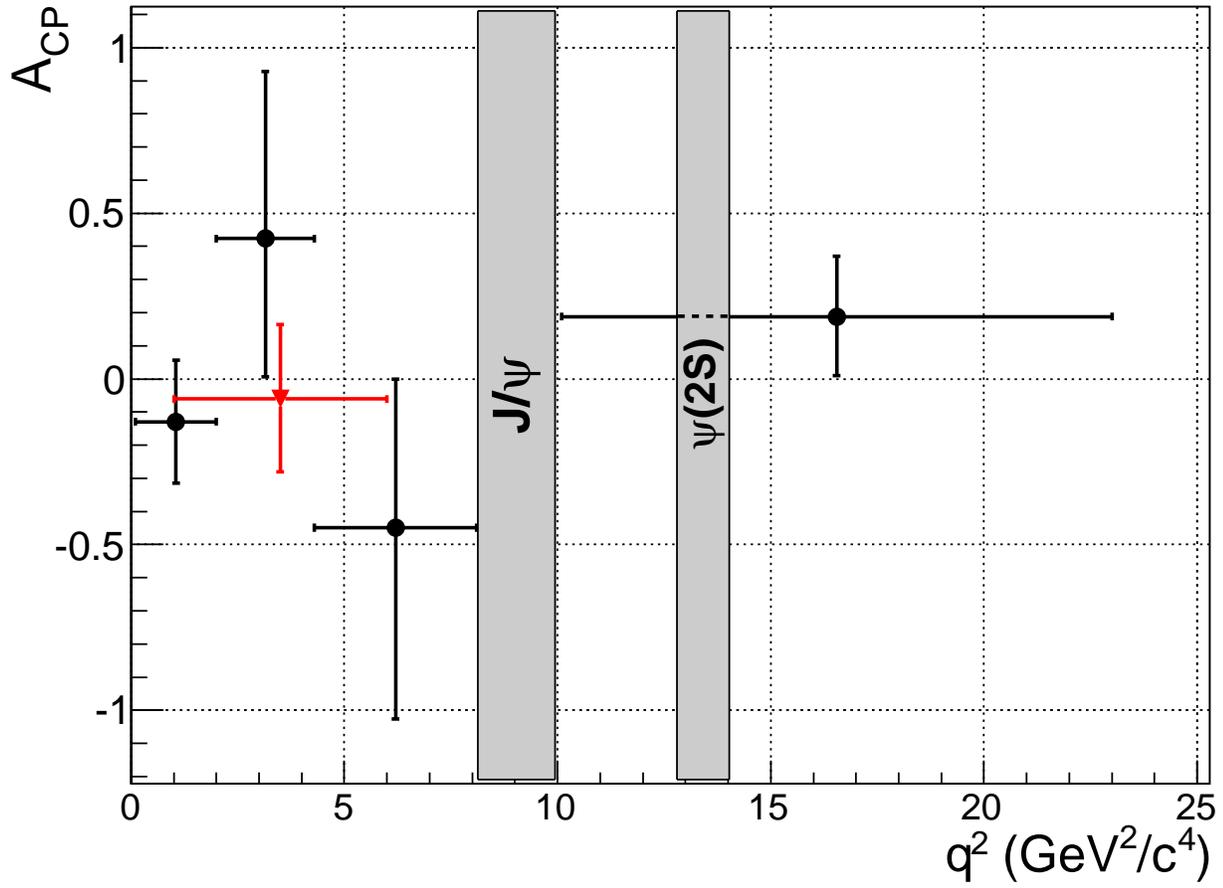}
  \caption{
    Results for $\acp$ as a function of $q^{2}$.
    The black points show the $q^{2}_{1}-q^{2}_{45}$ results; the red
    triangle denotes $q^{2}_{0}$. The $q^{2}_{45}$ $\acp$ result does not
    include events in the $\psitwos$ veto window.
  }\label{fig:acpresults}
\end{figure}

\begin{figure}[!h]
  \centering
  \includegraphics[width=\columnwidth]{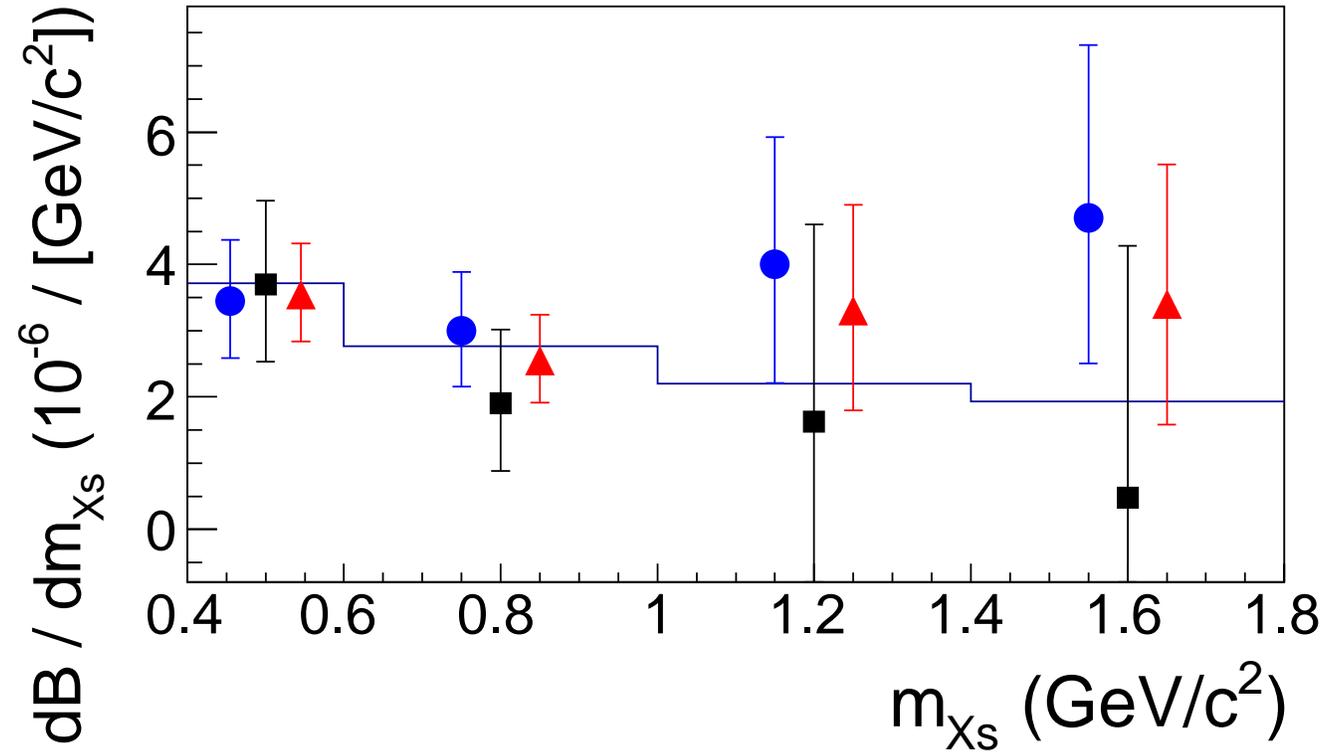}
  \caption{
    Differential BF as a function of $\mx$ for
    electron (blue circles), muon (black squares) and
    lepton-flavor-averaged final states (red triangles). The
    errors correspond to the total uncertainties.
    The histogram shows the SM expectation, which has
    uncertainties of approximately 10-30\% as a function of $q^{2}$.
    Estimated contributions from the vetoed charmonium $q^{2}$ regions are included.
    The horizontal spread of points in each bin is meant only to aid visibility.
  }\label{fig:mxresults}
\end{figure}

\begin{figure}[!htbp]
  \centering
  \includegraphics[width=\textwidth]{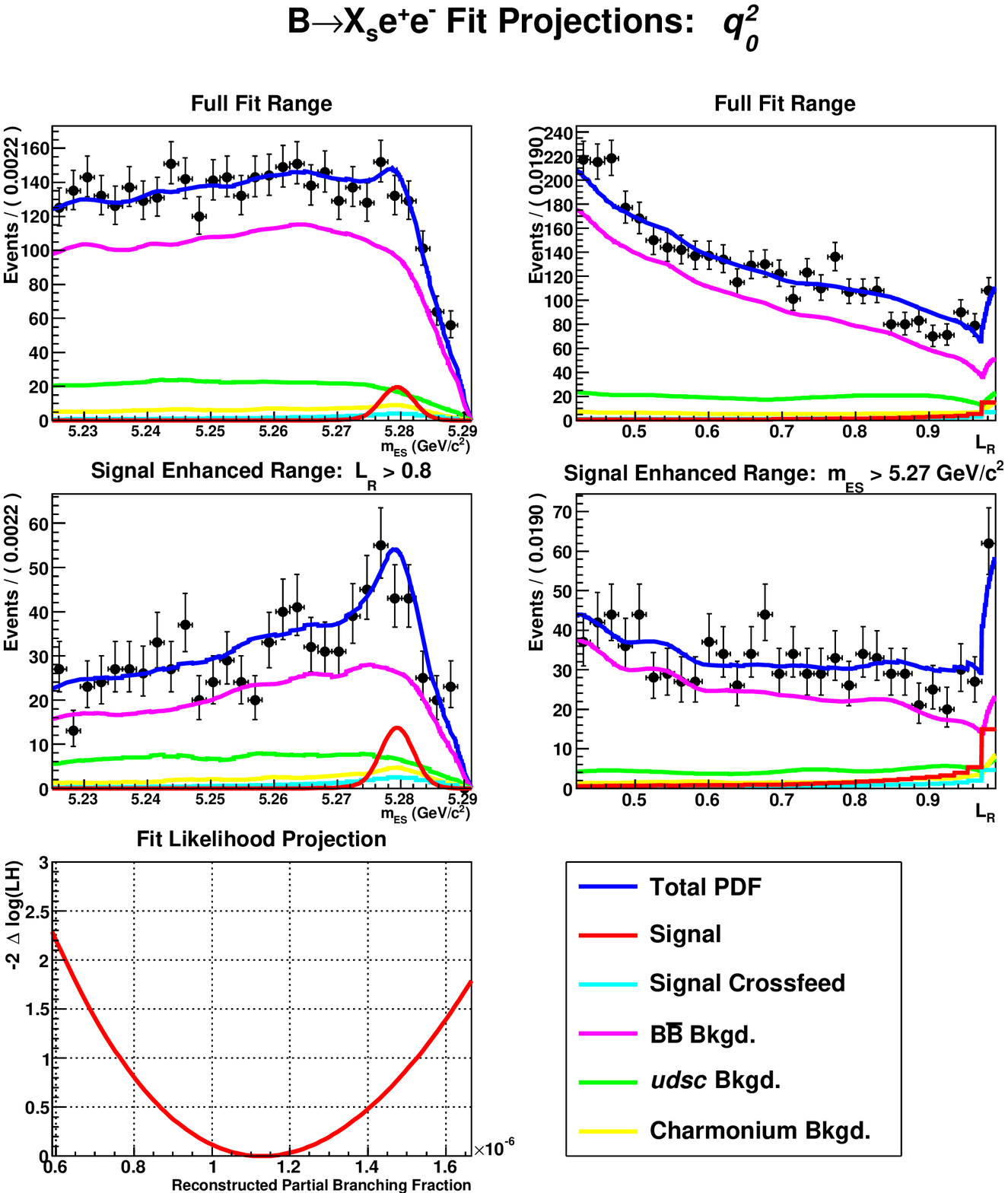}
  \caption[\BToXee\ qbin0 Fit Projection]{
    Fit to \BToXee\ in the $q^{2}_{0}$ bin.  Top row left is the $\mes$
    fit projection, top row right is the $\lhr$ fit projection; middle row left is a signal-enhanced
    $\mes$ fit projection for events with $\lhr > 0.8$, middle row right is a
    signal-enhanced $\lhr$ fit projection for events in the $\mes > 5.27 \gevcc$ signal region.
    The lower left hand plot is the profile likelihood curve for the 2D data fit.
  }\label{fig:ratesFit-ee-qbin0}
\end{figure}

\begin{figure}[!htbp]
  \centering
  \includegraphics[width=\textwidth]{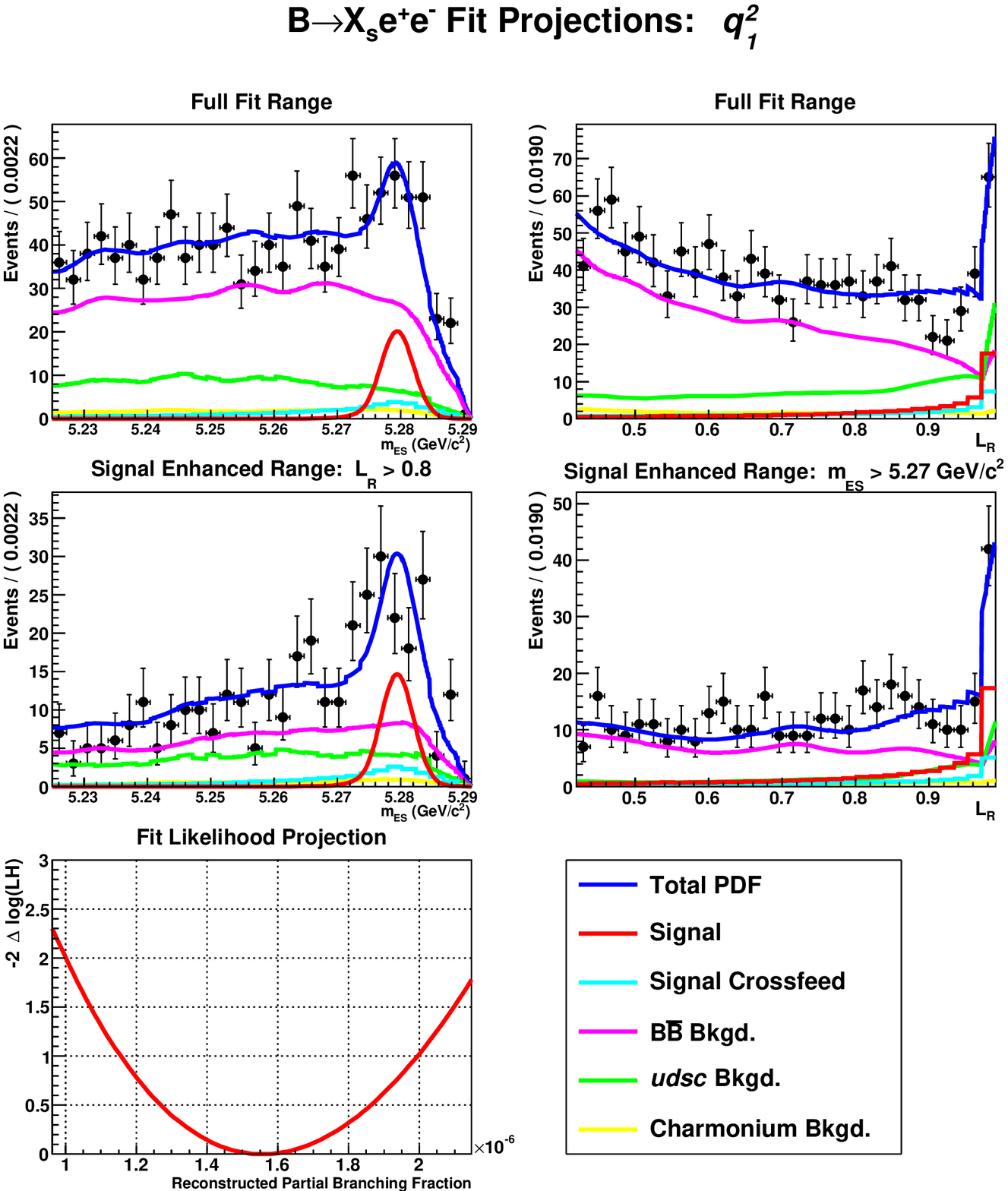}
  \caption[\BToXee\ qbin1 Fit Projection]{
    Fit to \BToXee\ in the $q^{2}_{1}$ bin.  Top row left is the $\mes$
    fit projection, top row right is the $\lhr$ fit projection; middle row left is a signal-enhanced
    $\mes$ fit projection for events with $\lhr > 0.8$, middle row right is a
    signal-enhanced $\lhr$ fit projection for events in the $\mes > 5.27 \gevcc$ signal region.
    The lower left hand plot is the profile likelihood curve for the 2D data fit.
  }\label{fig:ratesFit-ee-qbin1}
\end{figure}

\begin{figure}[!htbp]
  \centering
  \includegraphics[width=\textwidth]{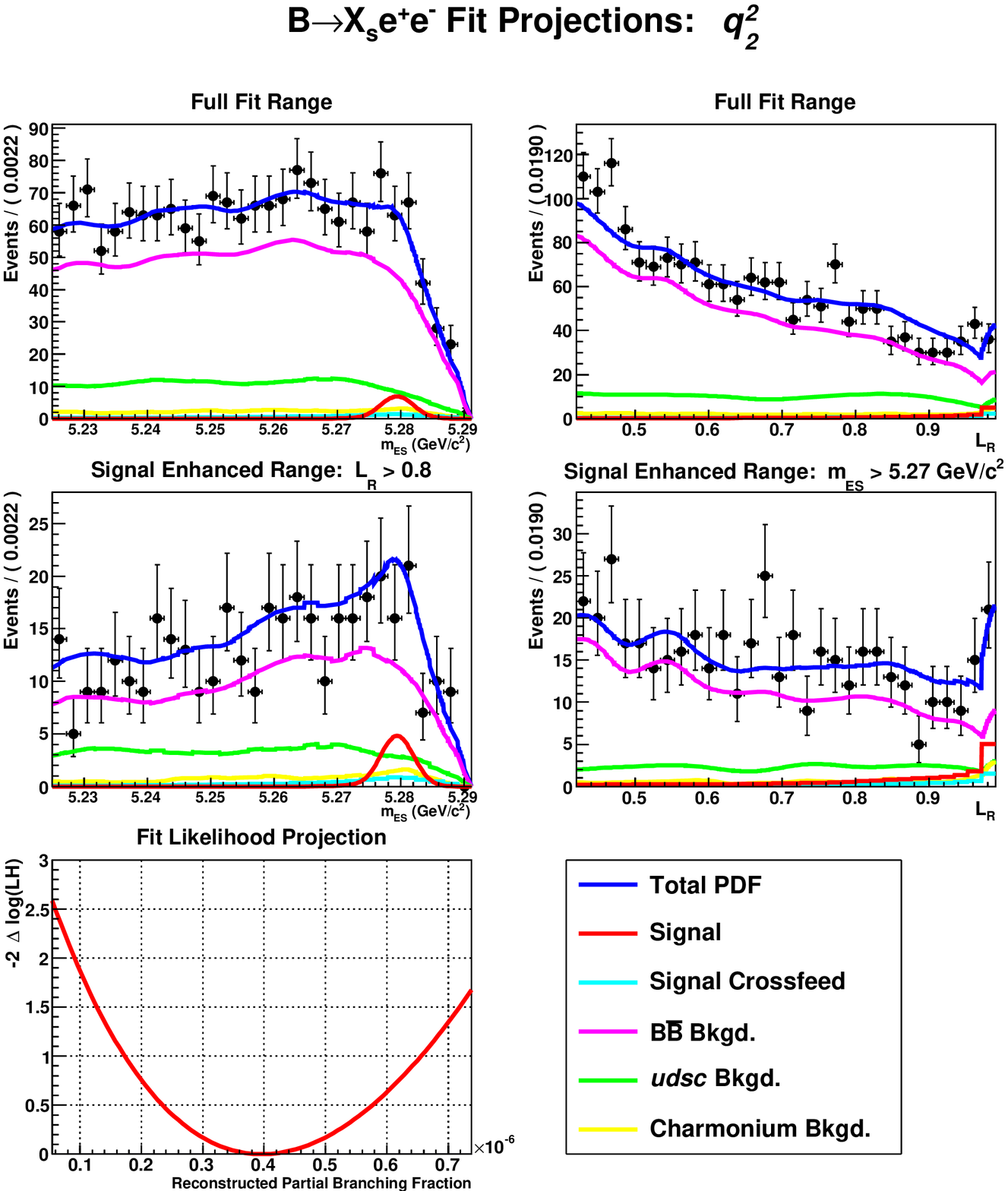}
  \caption[\BToXee\ qbin2 Fit Projection]{
    Fit to \BToXee\ in the $q^{2}_{2}$ bin.  Top row left is the $\mes$
    fit projection, top row right is the $\lhr$ fit projection; middle row left is a signal-enhanced
    $\mes$ fit projection for events with $\lhr > 0.8$, middle row right is a
    signal-enhanced $\lhr$ fit projection for events in the $\mes > 5.27 \gevcc$ signal region.
    The lower left hand plot is the profile likelihood curve for the 2D data fit.
  }\label{fig:ratesFit-ee-qbin2}
\end{figure}

\begin{figure}[!htbp]
  \centering
  \includegraphics[width=\textwidth]{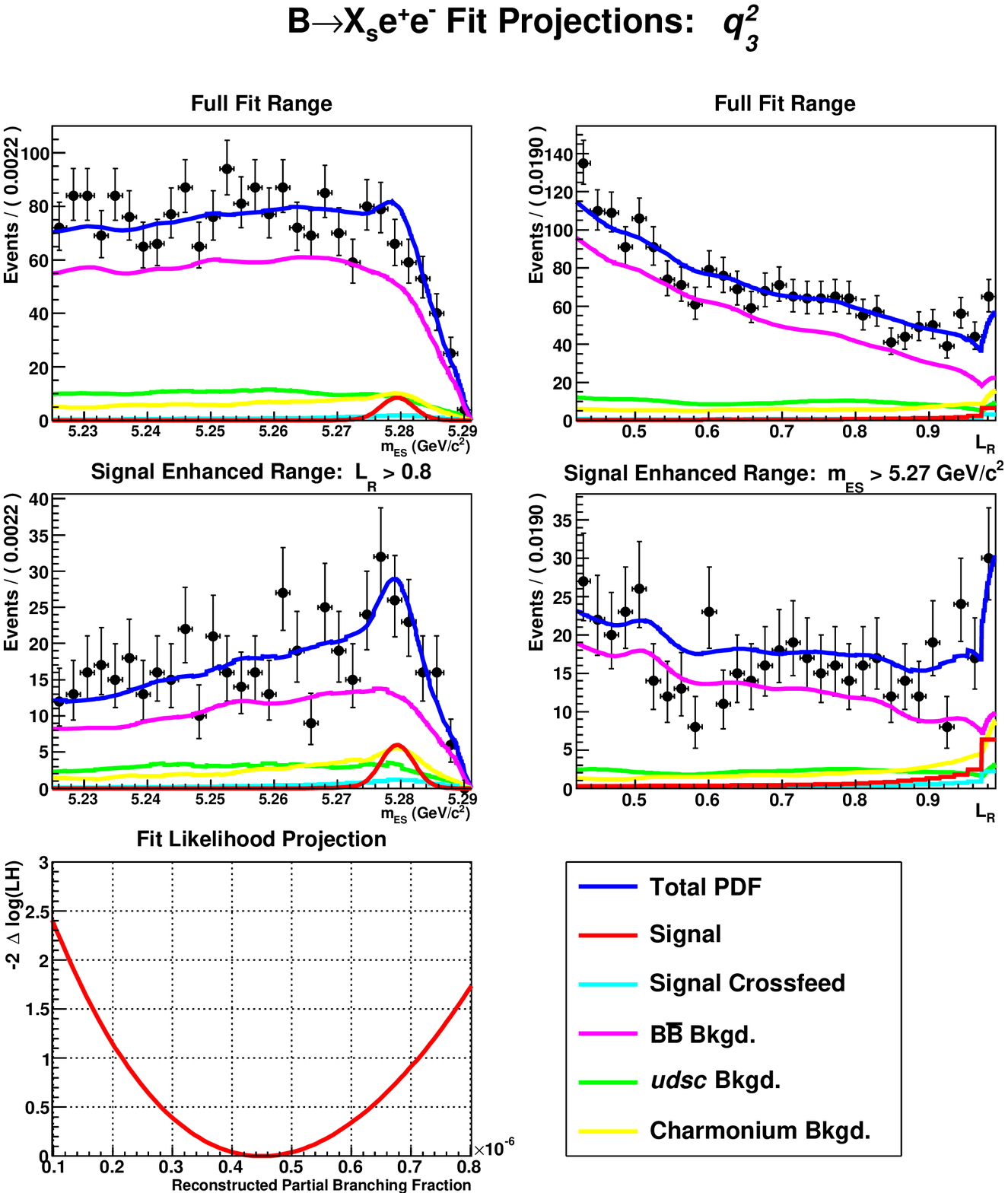}
  \caption[\BToXee\ qbin3 Fit Projection]{
    Fit to \BToXee\ in the $q^{2}_{3}$ bin.  Top row left is the $\mes$
    fit projection, top row right is the $\lhr$ fit projection; middle row left is a signal-enhanced
    $\mes$ fit projection for events with $\lhr > 0.8$, middle row right is a
    signal-enhanced $\lhr$ fit projection for events in the $\mes > 5.27 \gevcc$ signal region.
    The lower left hand plot is the profile likelihood curve for the 2D data fit.
  }\label{fig:ratesFit-ee-qbin3}
\end{figure}

\begin{figure}[!htbp]
  \centering
  \includegraphics[width=\textwidth]{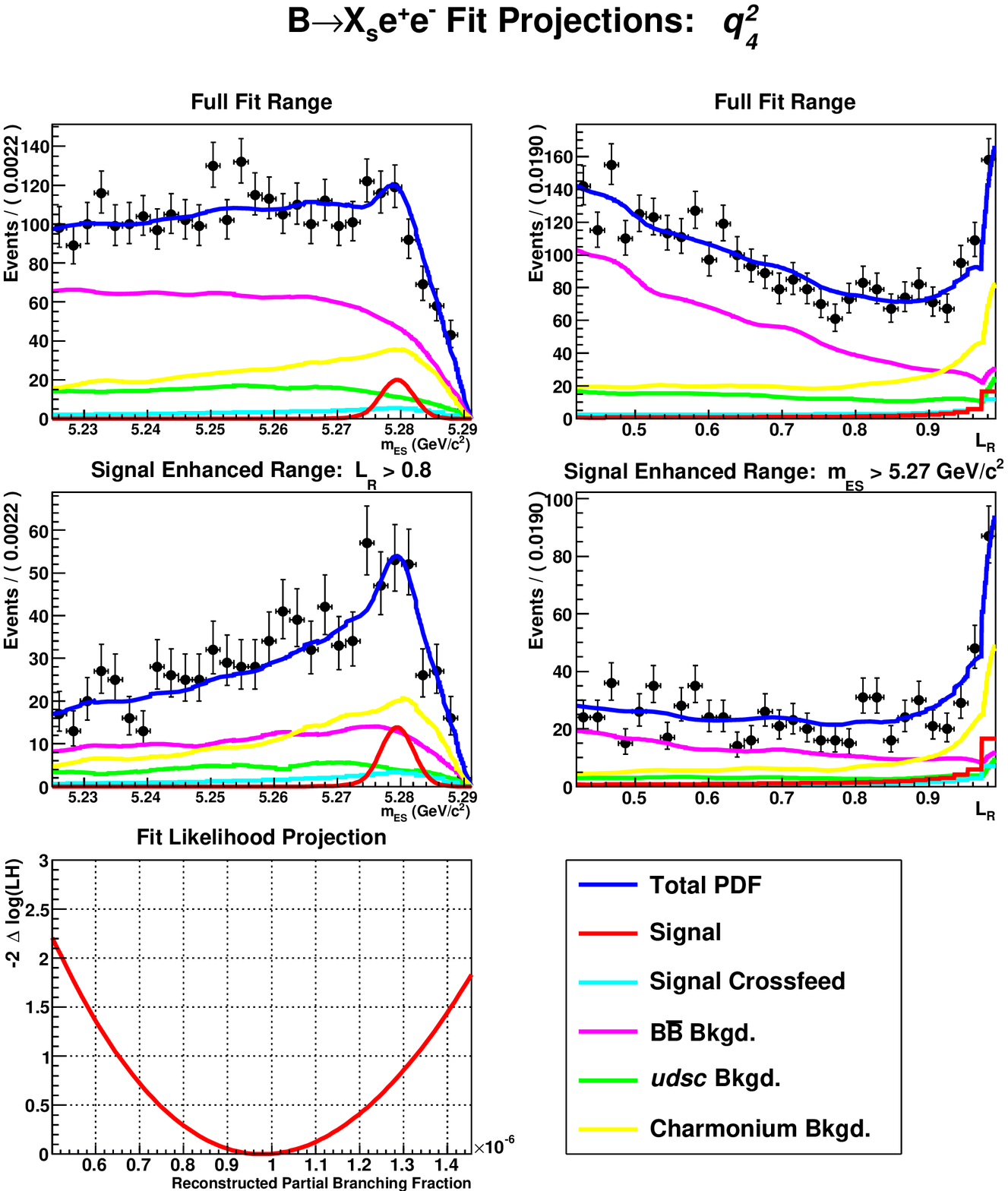}
  \caption[\BToXee\ qbin4 Fit Projection]{
    Fit to \BToXee\ in the $q^{2}_{4}$ bin.  Top row left is the $\mes$
    fit projection, top row right is the $\lhr$ fit projection; middle row left is a signal-enhanced
    $\mes$ fit projection for events with $\lhr > 0.8$, middle row right is a
    signal-enhanced $\lhr$ fit projection for events in the $\mes > 5.27 \gevcc$ signal region.
    The lower left hand plot is the profile likelihood curve for the 2D data fit.
  }\label{fig:ratesFit-ee-qbin4}
\end{figure}

\begin{figure}[!htbp]
  \centering
  \includegraphics[width=\textwidth]{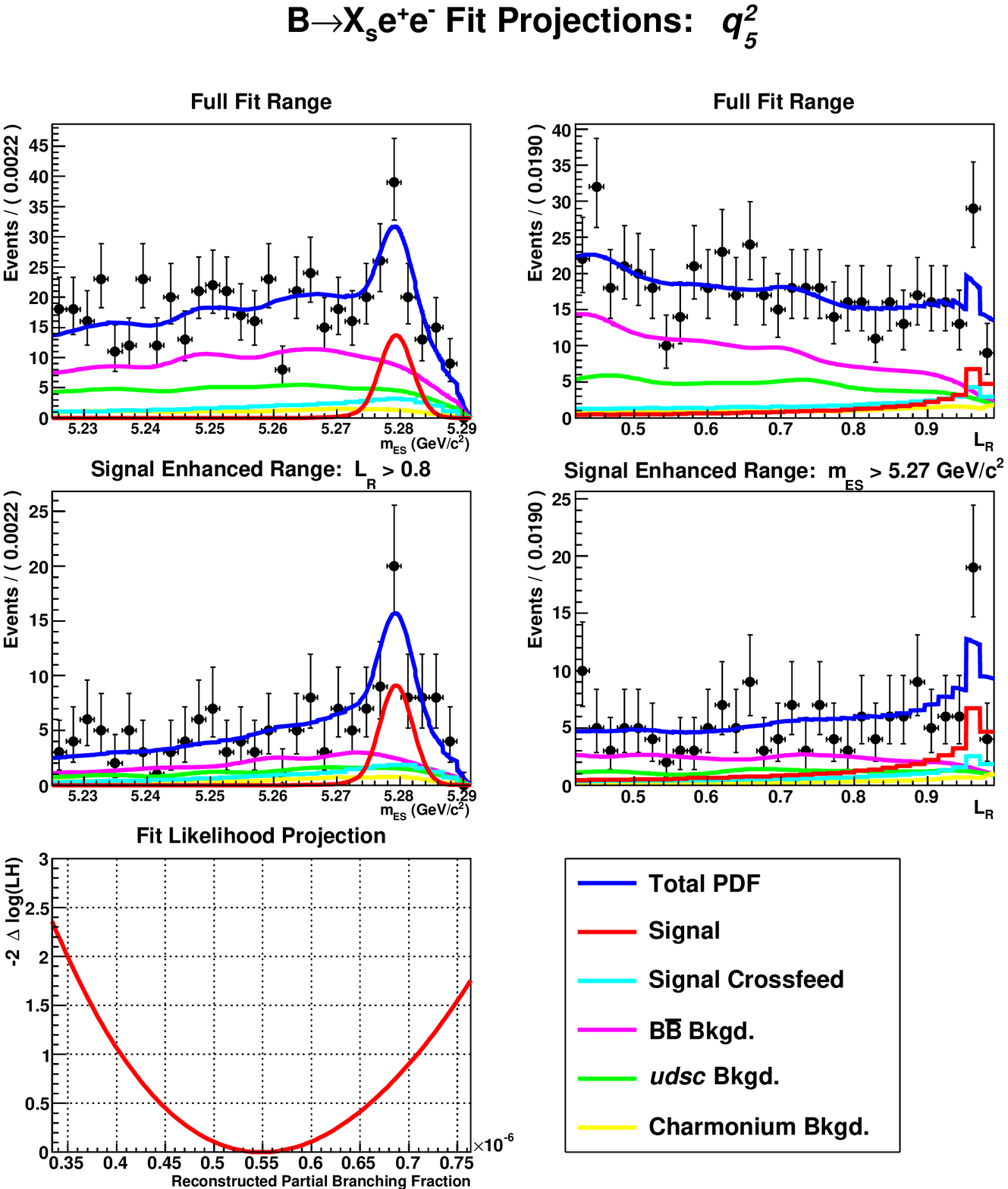}
  \caption[\BToXee\ qbin5 Fit Projection]{
    Fit to \BToXee\ in the $q^{2}_{5}$ bin.  Top row left is the $\mes$
    fit projection, top row right is the $\lhr$ fit projection; middle row left is a signal-enhanced
    $\mes$ fit projection for events with $\lhr > 0.8$, middle row right is a
    signal-enhanced $\lhr$ fit projection for events in the $\mes > 5.27 \gevcc$ signal region.
    The lower left hand plot is the profile likelihood curve for the 2D data fit.
  }\label{fig:ratesFit-ee-qbin5}
\end{figure}

\begin{figure}[!htbp]
  \centering
  \includegraphics[width=\textwidth]{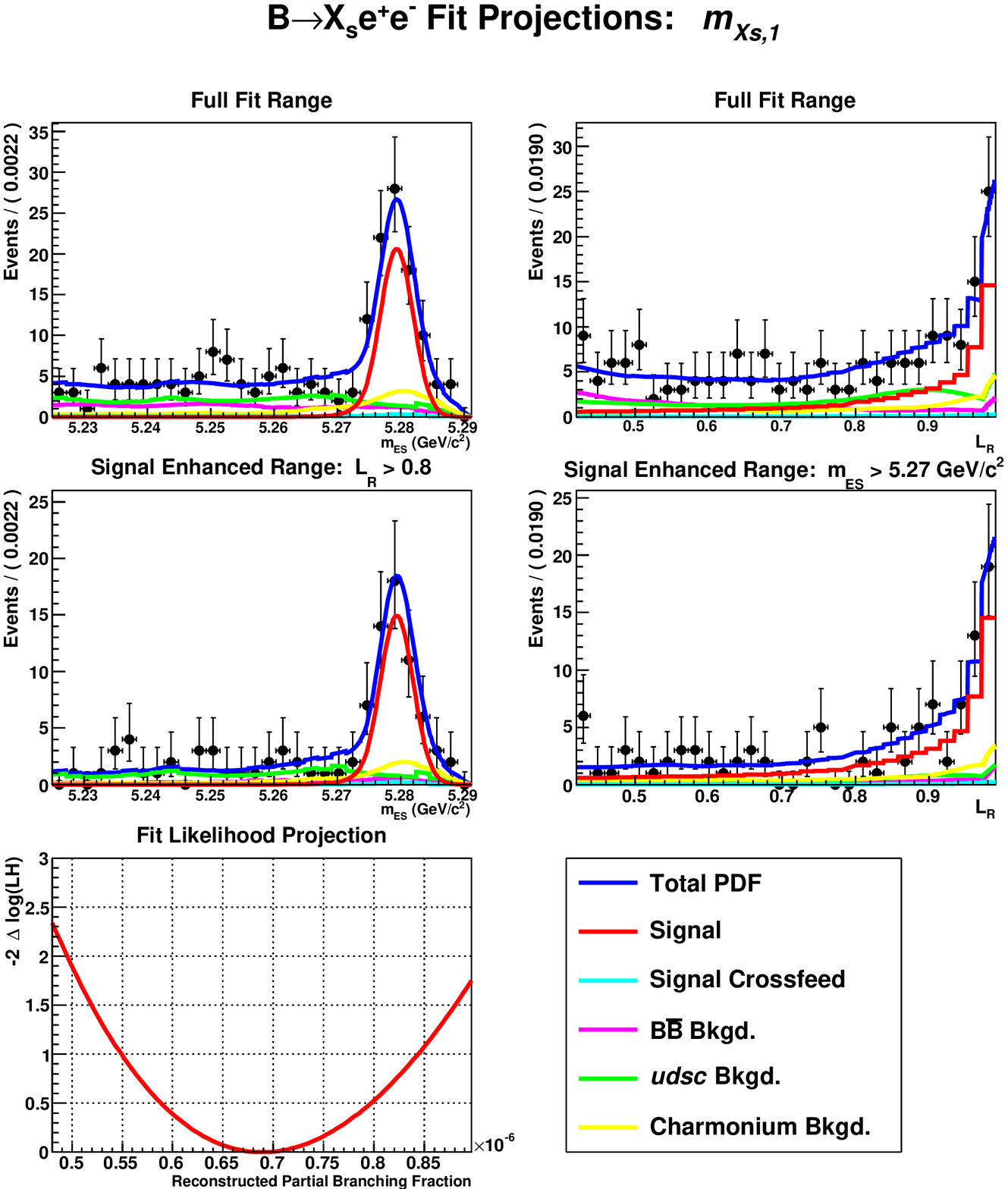}
  \caption[\BToXee\ mhad1 Fit Projection]{
    Fit to \BToXee\ in the $\mxone$ bin.  Top row left is the $\mes$
    fit projection, top row right is the $\lhr$ fit projection; middle row left is a signal-enhanced
    $\mes$ fit projection for events with $\lhr > 0.8$, middle row right is a
    signal-enhanced $\lhr$ fit projection for events in the $\mes > 5.27 \gevcc$ signal region.
    The lower left hand plot is the profile likelihood curve for the 2D data fit.
  }\label{fig:ratesFit-ee-mhad1}
\end{figure}

\begin{figure}[!htbp]
  \centering
  \includegraphics[width=\textwidth]{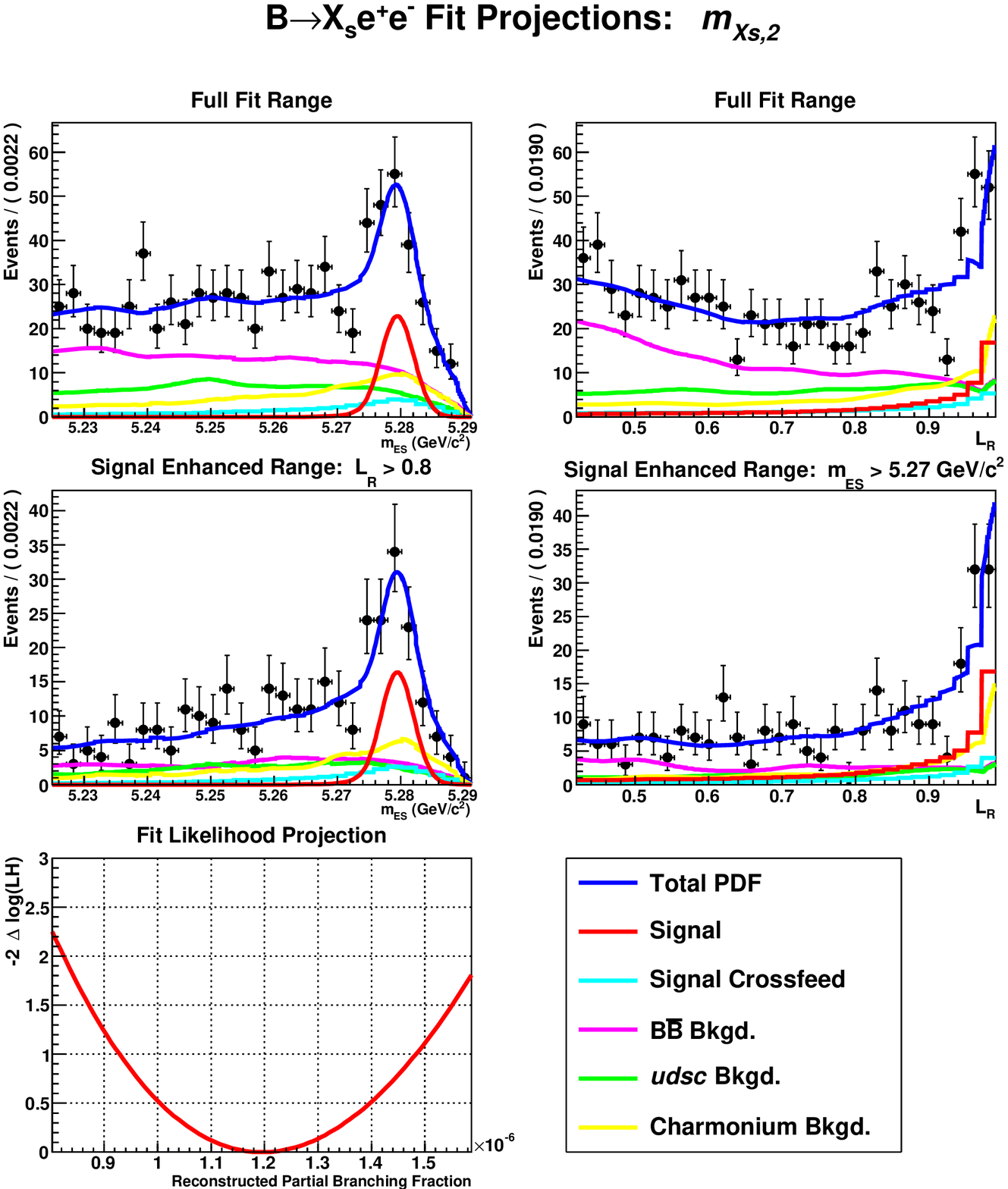}
  \caption[\BToXee\ mhad2 Fit Projection]{
    Fit to \BToXee\ in the $\mxtwo$ bin.  Top row left is the $\mes$
    fit projection, top row right is the $\lhr$ fit projection; middle row left is a signal-enhanced
    $\mes$ fit projection for events with $\lhr > 0.8$, middle row right is a
    signal-enhanced $\lhr$ fit projection for events in the $\mes > 5.27 \gevcc$ signal region.
    The lower left hand plot is the profile likelihood curve for the 2D data fit.
  }\label{fig:ratesFit-ee-mhad2}
\end{figure}

\begin{figure}[!htbp]
  \centering
  \includegraphics[width=\textwidth]{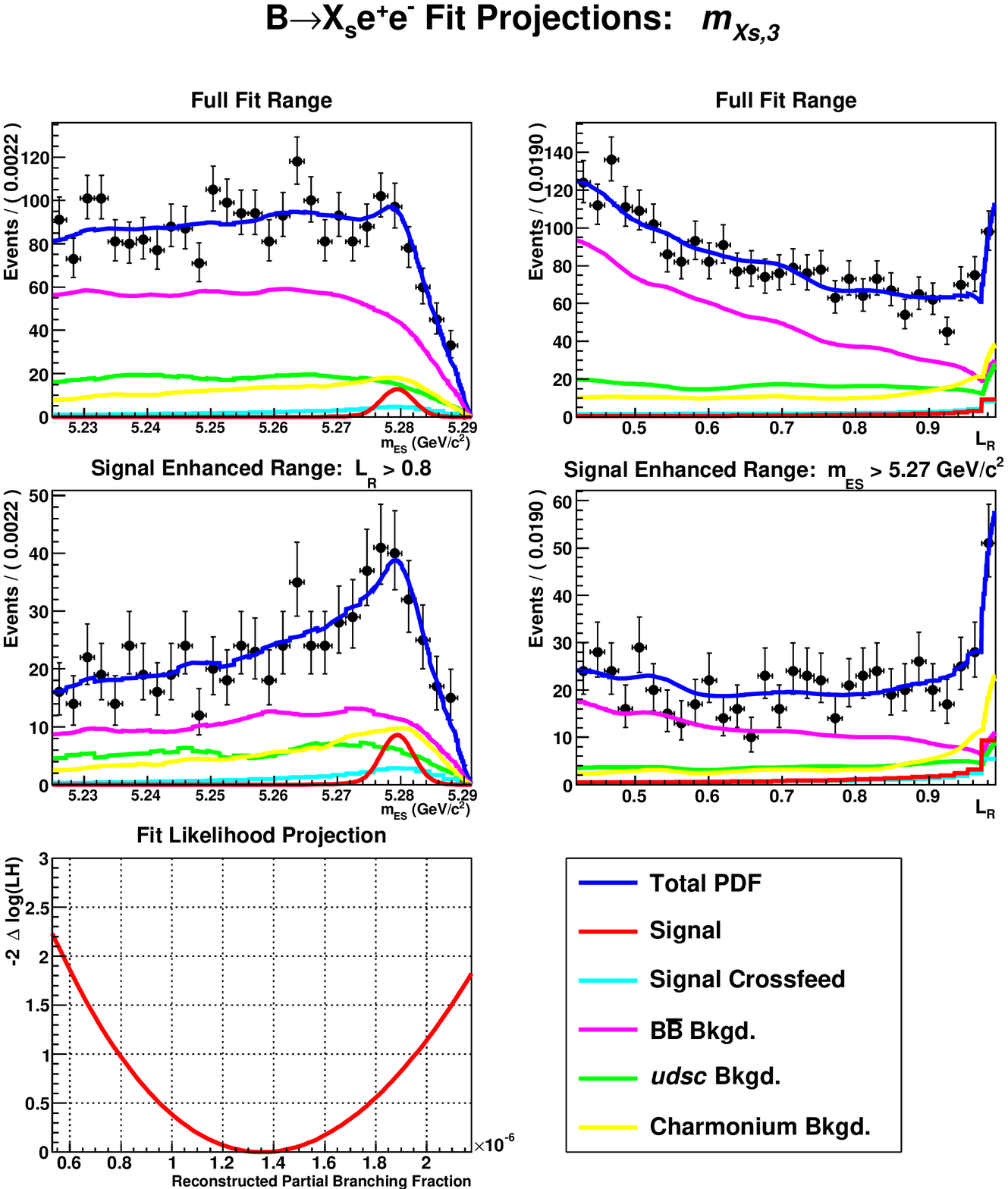}
  \caption[\BToXee\ mhad3 Fit Projection]{
    Fit to \BToXee\ in the $\mxthree$ bin.  Top row left is the $\mes$
    fit projection, top row right is the $\lhr$ fit projection; middle row left is a signal-enhanced
    $\mes$ fit projection for events with $\lhr > 0.8$, middle row right is a
    signal-enhanced $\lhr$ fit projection for events in the $\mes > 5.27 \gevcc$ signal region.
    The lower left hand plot is the profile likelihood curve for the 2D data fit.
  }\label{fig:ratesFit-ee-mhad3}
\end{figure}

\begin{figure}[!htbp]
  \centering
  \includegraphics[width=\textwidth]{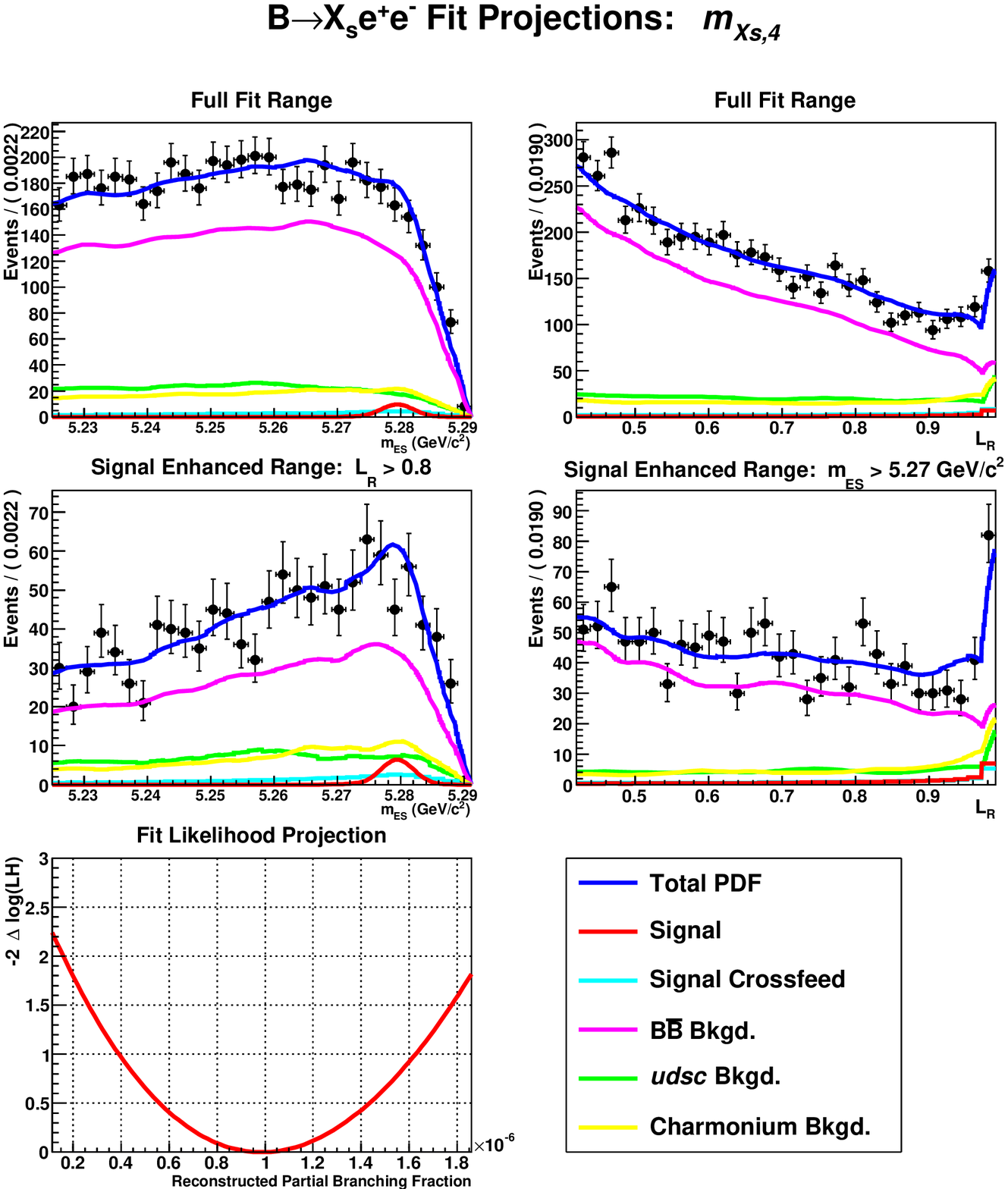}
  \caption[\BToXee\ mhad4 Fit Projection]{
    Fit to \BToXee\ in the $\mxfour$ bin.  Top row left is the $\mes$
    fit projection, top row right is the $\lhr$ fit projection; middle row left is a signal-enhanced
    $\mes$ fit projection for events with $\lhr > 0.8$, middle row right is a
    signal-enhanced $\lhr$ fit projection for events in the $\mes > 5.27 \gevcc$ signal region.
    The lower left hand plot is the profile likelihood curve for the 2D data fit.
  }\label{fig:ratesFit-ee-mhad4}
\end{figure}

\begin{figure}[!htbp]
  \centering
  \includegraphics[width=\textwidth]{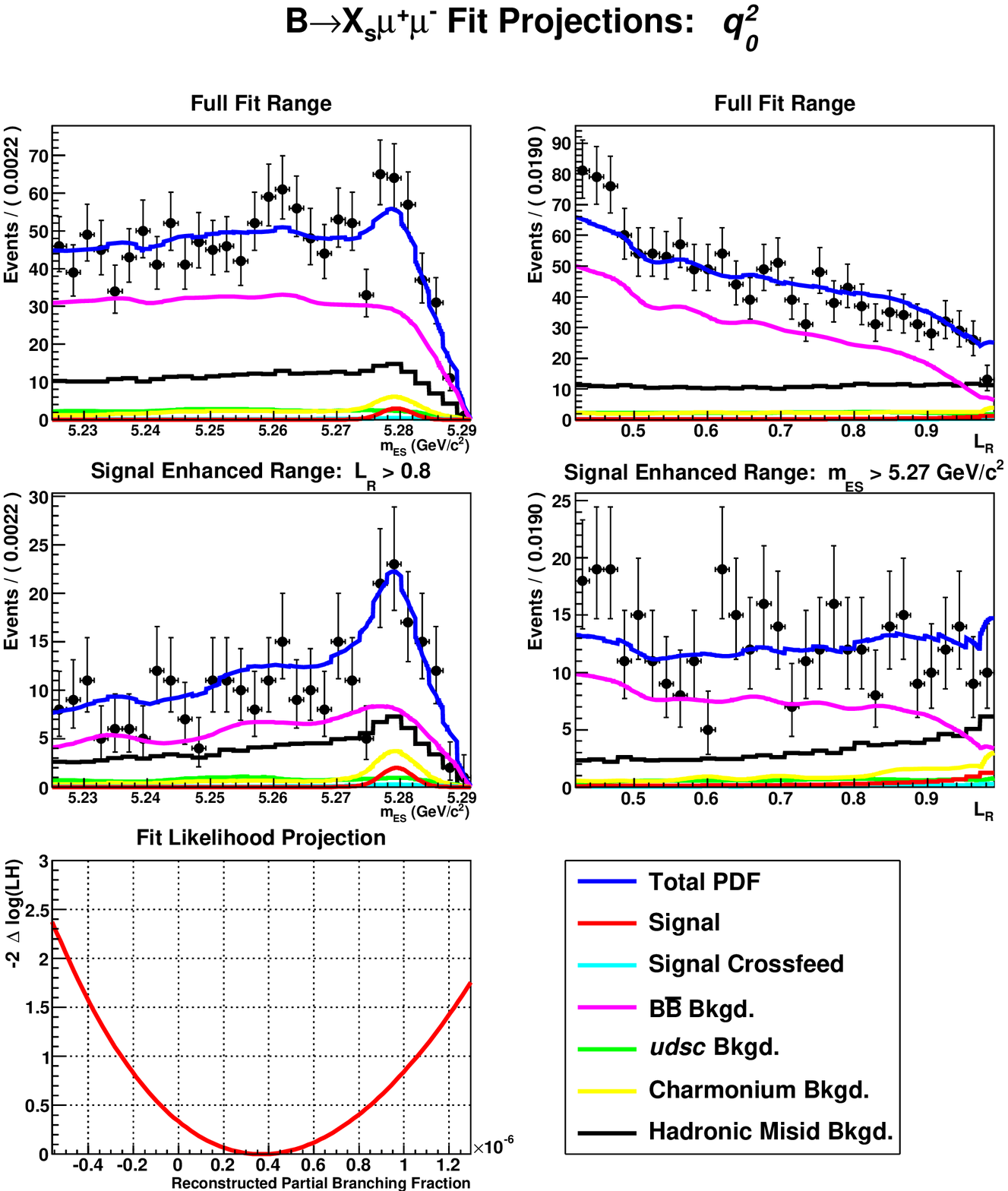}
  \caption[\BToXmm\ qbin0 Fit Projection]{
    Fit to \BToXmm\ in the $q^{2}_{0}$ bin.  Top row left is the $\mes$
    fit projection, top row right is the $\lhr$ fit projection; middle row left is a signal-enhanced
    $\mes$ fit projection for events with $\lhr > 0.8$, middle row right is a
    signal-enhanced $\lhr$ fit projection for events in the $\mes > 5.27 \gevcc$ signal region.
    The lower left hand plot is the profile likelihood curve for the 2D data fit.
  }\label{fig:ratesFit-mm-qbin0}
\end{figure}

\begin{figure}[!htbp]
  \centering
  \includegraphics[width=\textwidth]{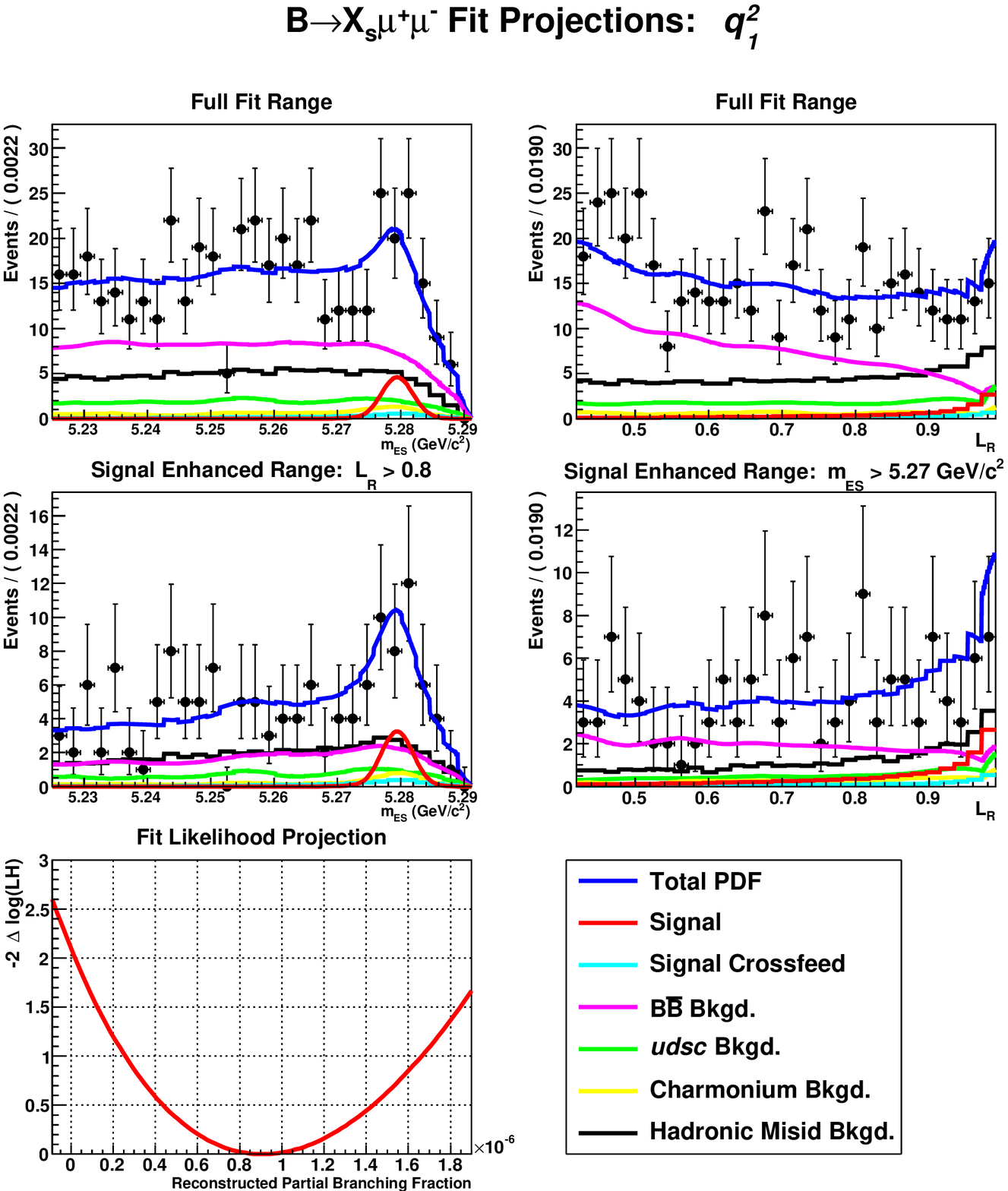}
  \caption[\BToXmm\ qbin1 Fit Projection]{
    Fit to \BToXmm\ in the $q^{2}_{1}$ bin.  Top row left is the $\mes$
    fit projection, top row right is the $\lhr$ fit projection; middle row left is a signal-enhanced
    $\mes$ fit projection for events with $\lhr > 0.8$, middle row right is a
    signal-enhanced $\lhr$ fit projection for events in the $\mes > 5.27 \gevcc$ signal region.
    The lower left hand plot is the profile likelihood curve for the 2D data fit.
  }\label{fig:ratesFit-mm-qbin1}
\end{figure}

\begin{figure}[!htbp]
  \centering
  \includegraphics[width=\textwidth]{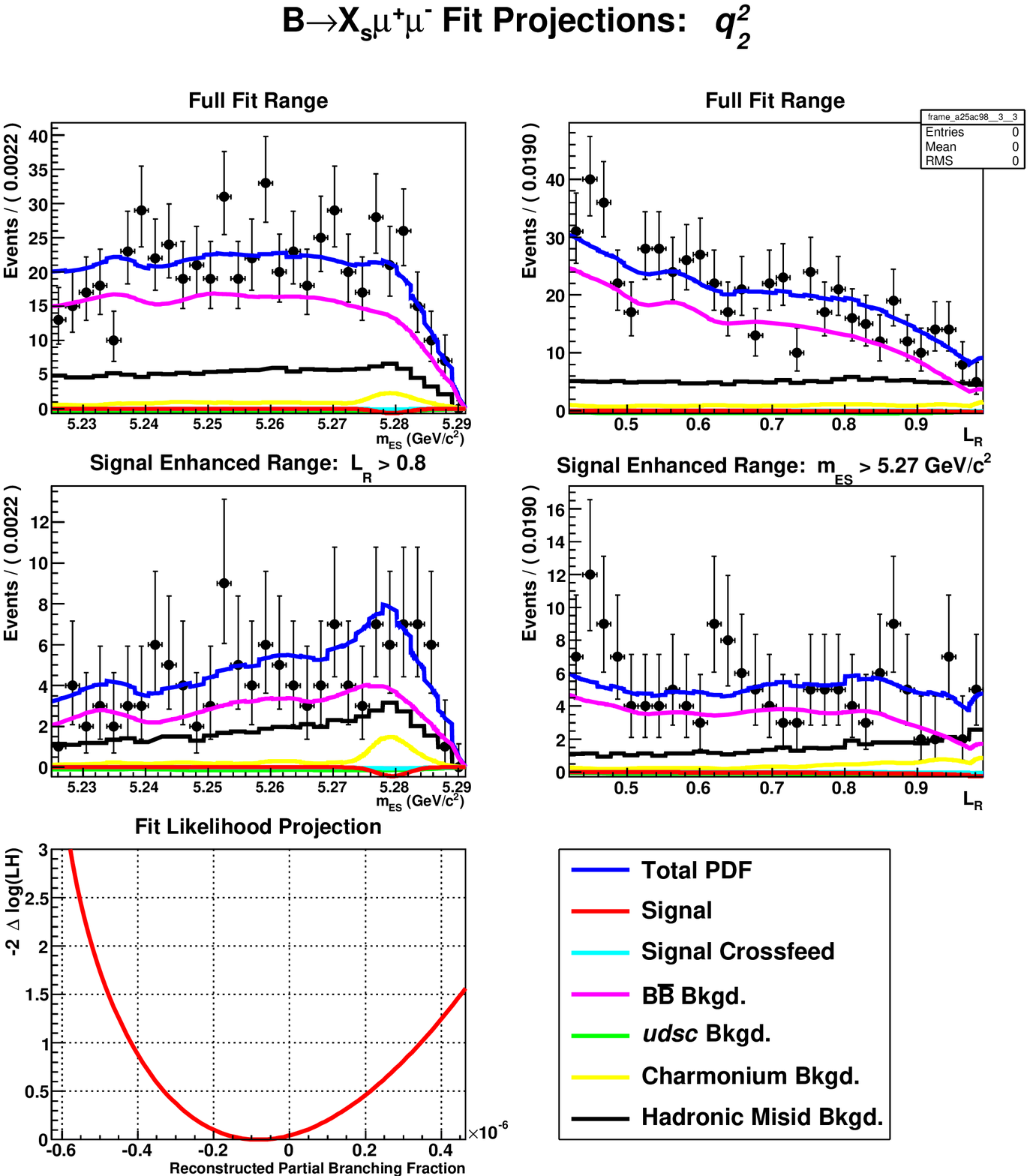}
  \caption[\BToXmm\ qbin2 Fit Projection]{
    Fit to \BToXmm\ in the $q^{2}_{2}$ bin.  Top row left is the $\mes$
    fit projection, top row right is the $\lhr$ fit projection; middle row left is a signal-enhanced
    $\mes$ fit projection for events with $\lhr > 0.8$, middle row right is a
    signal-enhanced $\lhr$ fit projection for events in the $\mes > 5.27 \gevcc$ signal region.
    The lower left hand plot is the profile likelihood curve for the 2D data fit.
  }\label{fig:ratesFit-mm-qbin2}
\end{figure}

\begin{figure}[!htbp]
  \centering
  \includegraphics[width=\textwidth]{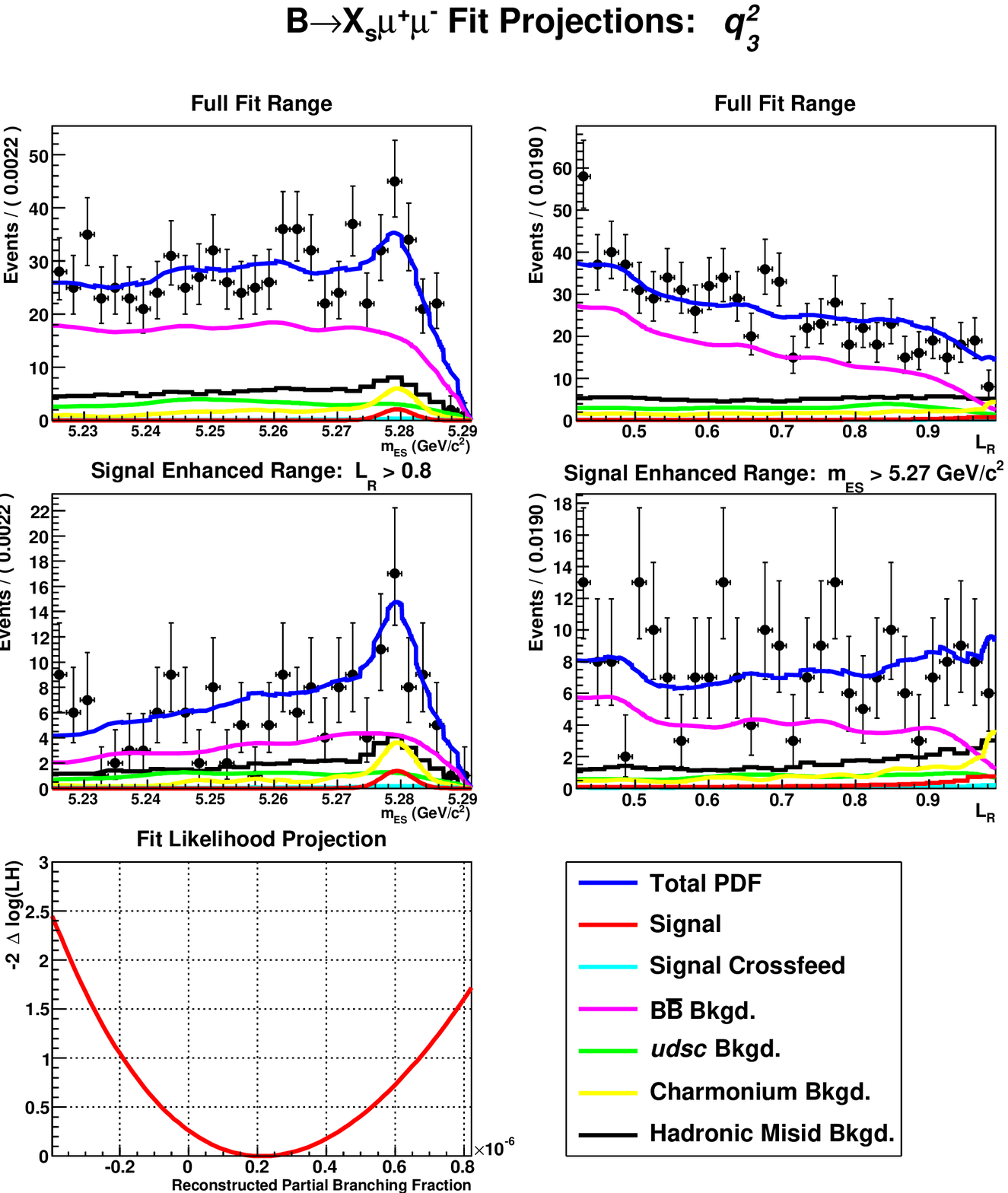}
  \caption[\BToXmm\ qbin3 Fit Projection]{
    Fit to \BToXmm\ in the $q^{2}_{3}$ bin.  Top row left is the $\mes$
    fit projection, top row right is the $\lhr$ fit projection; middle row left is a signal-enhanced
    $\mes$ fit projection for events with $\lhr > 0.8$, middle row right is a
    signal-enhanced $\lhr$ fit projection for events in the $\mes > 5.27 \gevcc$ signal region.
    The lower left hand plot is the profile likelihood curve for the 2D data fit.
  }\label{fig:ratesFit-mm-qbin3}
\end{figure}

\begin{figure}[!htbp]
  \centering
  \includegraphics[width=\textwidth]{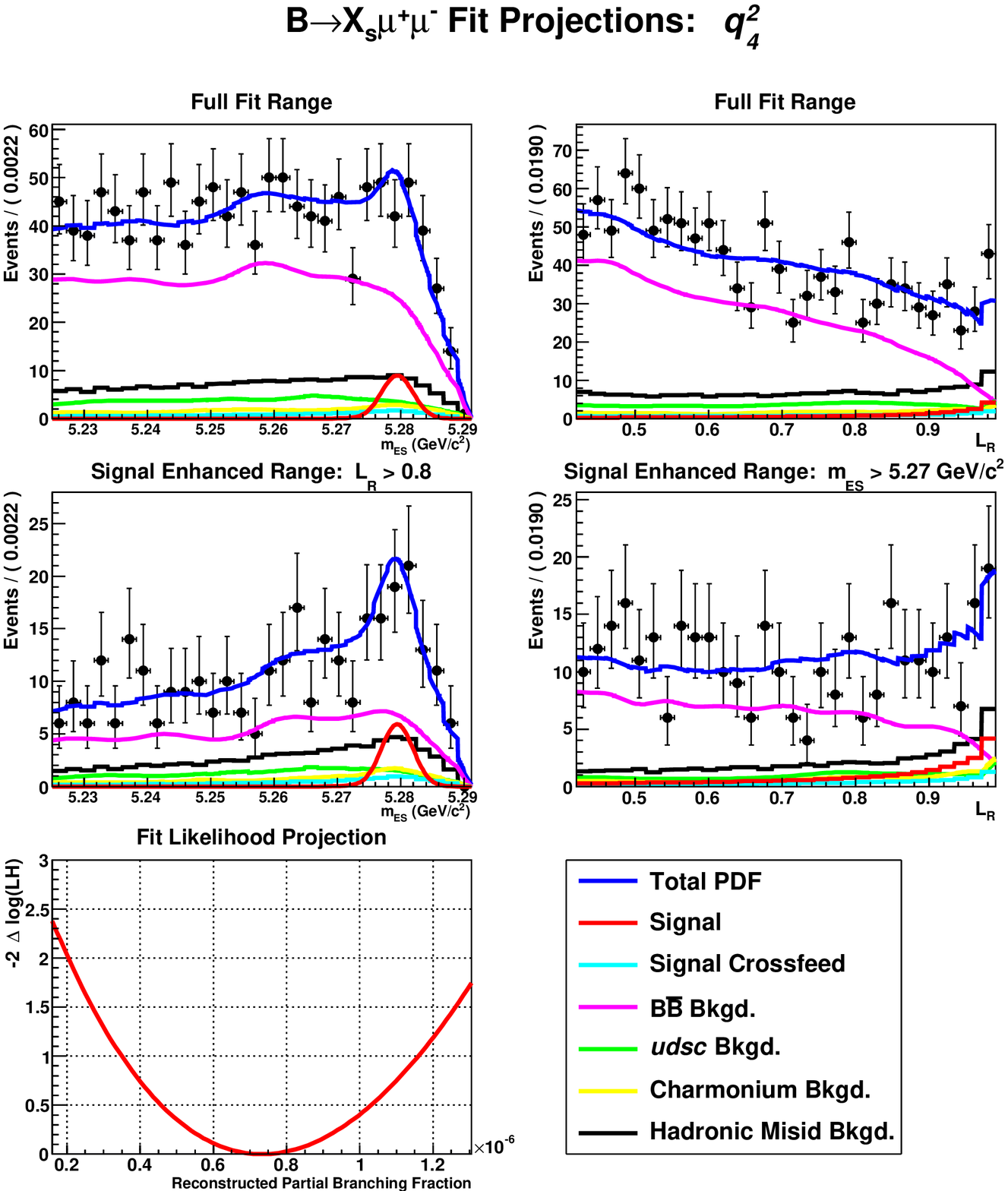}
  \caption[\BToXmm\ qbin4 Fit Projection]{
    Fit to \BToXmm\ in the $q^{2}_{4}$ bin.  Top row left is the $\mes$
    fit projection, top row right is the $\lhr$ fit projection; middle row left is a signal-enhanced
    $\mes$ fit projection for events with $\lhr > 0.8$, middle row right is a
    signal-enhanced $\lhr$ fit projection for events in the $\mes > 5.27 \gevcc$ signal region.
    The lower left hand plot is the profile likelihood curve for the 2D data fit.
  }\label{fig:ratesFit-mm-qbin4}
\end{figure}

\begin{figure}[!htbp]
  \centering
  \includegraphics[width=\textwidth]{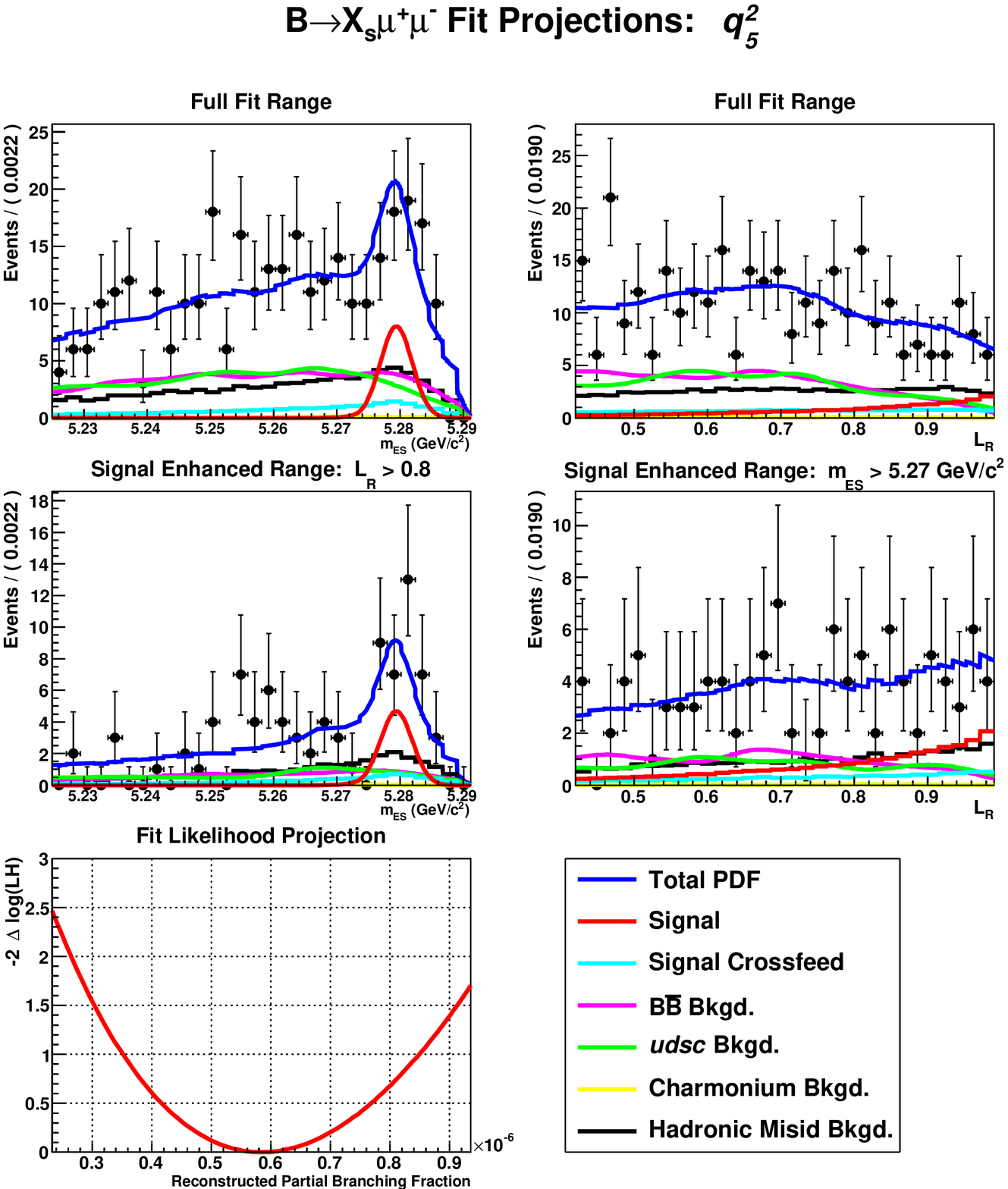}
  \caption[\BToXmm\ qbin5 Fit Projection]{
    Fit to \BToXmm\ in the $q^{2}_{5}$ bin.  Top row left is the $\mes$
    fit projection, top row right is the $\lhr$ fit projection; middle row left is a signal-enhanced
    $\mes$ fit projection for events with $\lhr > 0.8$, middle row right is a
    signal-enhanced $\lhr$ fit projection for events in the $\mes > 5.27 \gevcc$ signal region.
    The lower left hand plot is the profile likelihood curve for the 2D data fit.
  }\label{fig:ratesFit-mm-qbin5}
\end{figure}

\begin{figure}[!htbp]
  \centering
  \includegraphics[width=\textwidth]{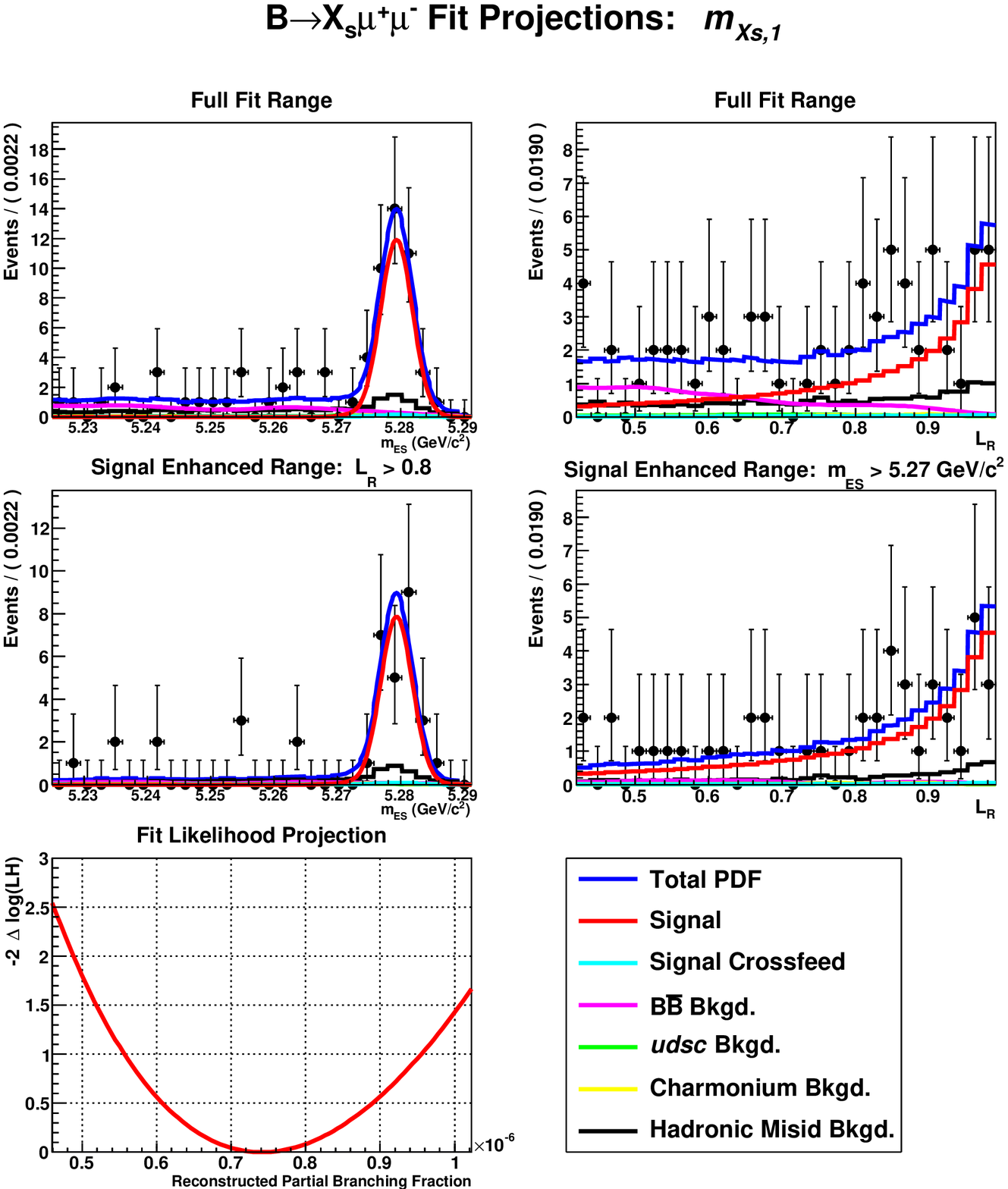}
  \caption[\BToXmm\ mhad1 Fit Projection]{
    Fit to \BToXmm\ in the $\mxone$ bin.  Top row left is the $\mes$
    fit projection, top row right is the $\lhr$ fit projection; middle row left is a signal-enhanced
    $\mes$ fit projection for events with $\lhr > 0.8$, middle row right is a
    signal-enhanced $\lhr$ fit projection for events in the $\mes > 5.27 \gevcc$ signal region.
    The lower left hand plot is the profile likelihood curve for the 2D data fit.
  }\label{fig:ratesFit-mm-mhad1}
\end{figure}

\begin{figure}[!htbp]
  \centering
  \includegraphics[width=\textwidth]{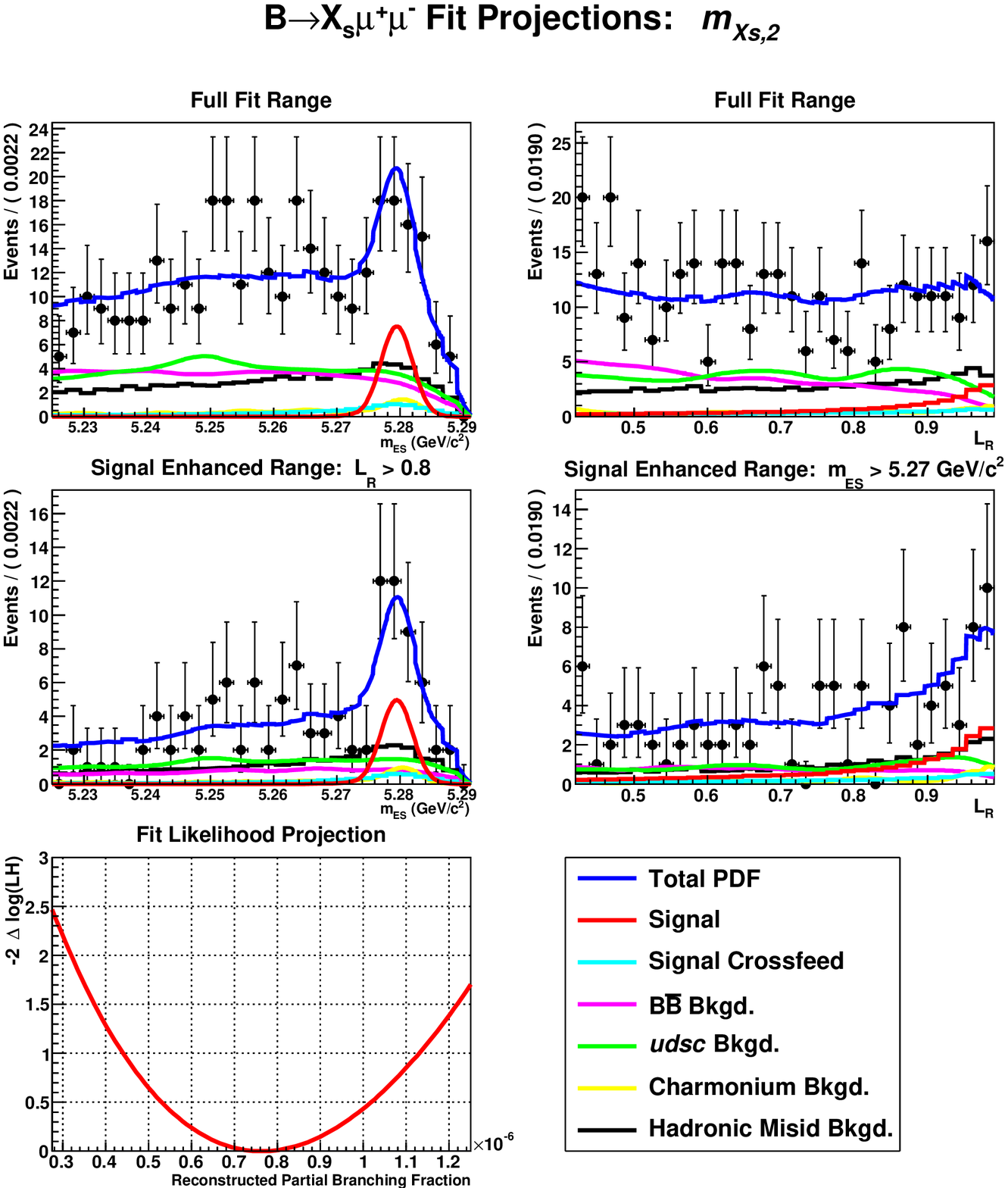}
  \caption[\BToXmm\ mhad2 Fit Projection]{
    Fit to \BToXmm\ in the $\mxtwo$ bin.  Top row left is the $\mes$
    fit projection, top row right is the $\lhr$ fit projection; middle row left is a signal-enhanced
    $\mes$ fit projection for events with $\lhr > 0.8$, middle row right is a
    signal-enhanced $\lhr$ fit projection for events in the $\mes > 5.27 \gevcc$ signal region.
    The lower left hand plot is the profile likelihood curve for the 2D data fit.
  }\label{fig:ratesFit-mm-mhad2}
\end{figure}

\begin{figure}[!htbp]
  \centering
  \includegraphics[width=\textwidth]{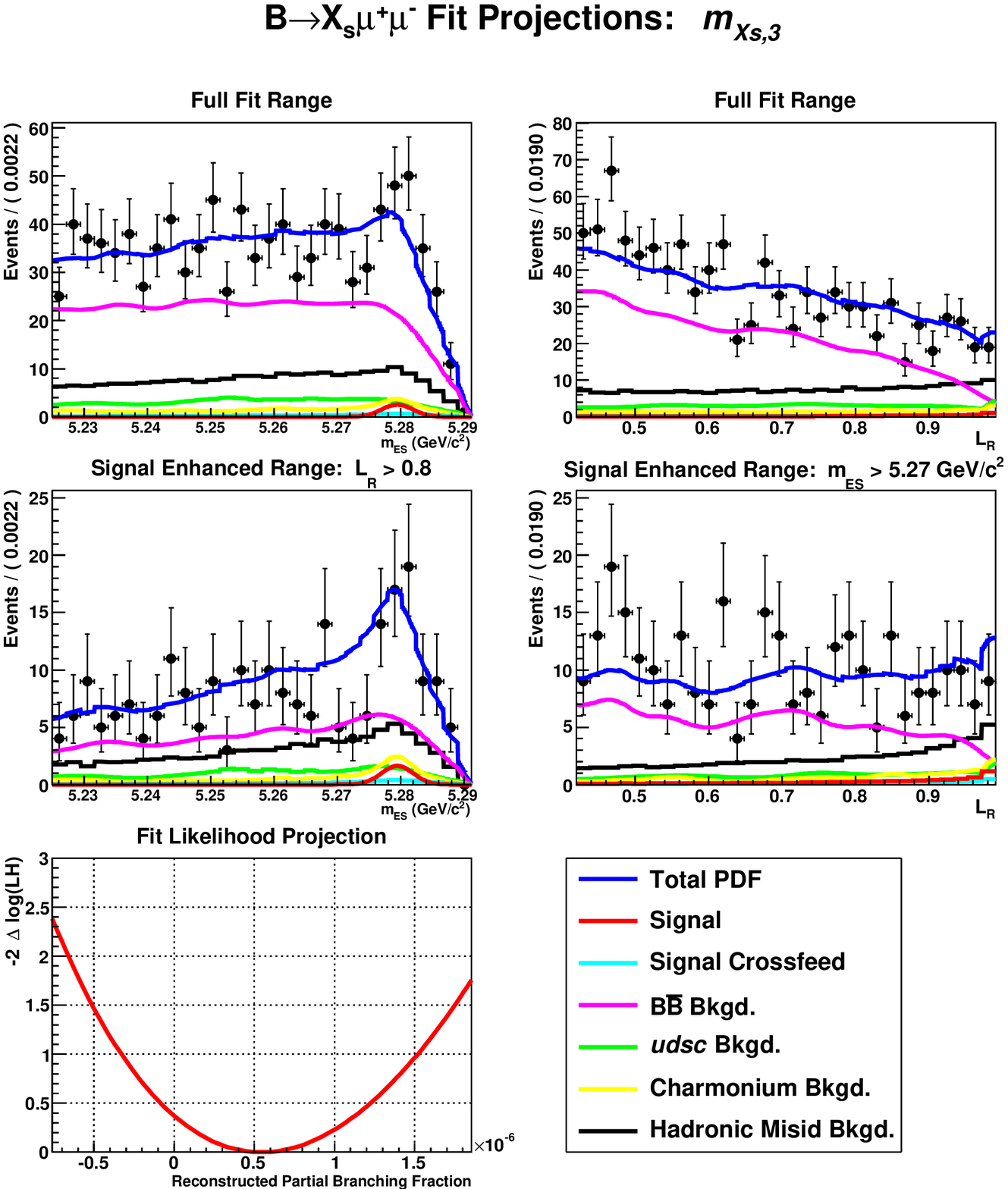}
  \caption[\BToXmm\ mhad3 Fit Projection]{
    Fit to \BToXmm\ in the $\mxthree$ bin.  Top row left is the $\mes$
    fit projection, top row right is the $\lhr$ fit projection; middle row left is a signal-enhanced
    $\mes$ fit projection for events with $\lhr > 0.8$, middle row right is a
    signal-enhanced $\lhr$ fit projection for events in the $\mes > 5.27 \gevcc$ signal region.
    The lower left hand plot is the profile likelihood curve for the 2D data fit.
  }\label{fig:ratesFit-mm-mhad3}
\end{figure}

\begin{figure}[!htbp]
  \centering
  \includegraphics[width=\textwidth]{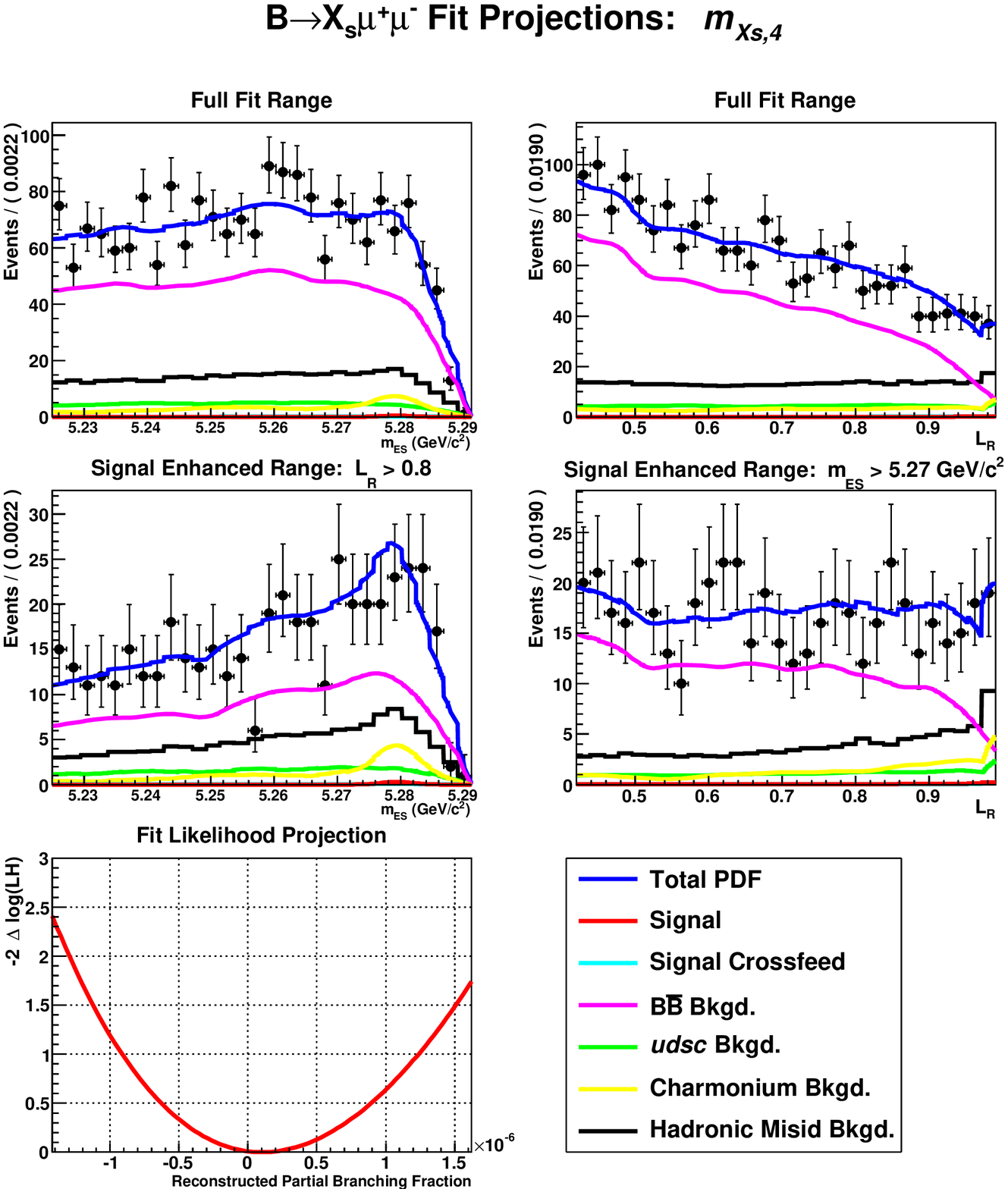}
  \caption[\BToXmm\ mhad4 Fit Projection]{
    Fit to \BToXmm\ in the $\mxfour$ bin.  Top row left is the $\mes$
    fit projection, top row right is the $\lhr$ fit projection; middle row left is a signal-enhanced
    $\mes$ fit projection for events with $\lhr > 0.8$, middle row right is a
    signal-enhanced $\lhr$ fit projection for events in the $\mes > 5.27 \gevcc$ signal region.
    The lower left hand plot is the profile likelihood curve for the 2D data fit.
  }\label{fig:ratesFit-mm-mhad4}
\end{figure}

This Supplemental Material includes:

\begin{itemize}
\item Figure~\ref{fig:acpresults}, plotting the $\acp$ results as a
function of $q^{2}$;

\item Figure~\ref{fig:mxresults}, plotting the differential branching fraction as a
function of $\mx$;

\item Table~\ref{tab:rawresults}, giving in each individual $q^{2}$ and $\mx$ bin
the fitted raw number of signal events $\nsig$, as well as the fitted number of
random combinatorial $\BB$ background events $\nbb$ present in the signal enhanced
region with $\mes > 5.27 \gevcc$;

\item Tables~\ref{tab:addsystrates-ee}-\ref{tab:modelsystrates-mm}, detailing
individual contributions to the ``additive'' and ``multiplicative'' branching
fraction systematics (as defined in the article main text), and the
model-dependent extrapolation systematics; and

\item Figures~\ref{fig:ratesFit-ee-qbin0}-\ref{fig:ratesFit-mm-mhad4},
which show the projections of our branching fraction fits onto their
respective datasets.

\end{itemize}

\newpage

\subsection{$\acp$ results.}

\FloatBarrier

\newpage

\subsection{Differental Branching Fraction in $\mx$ bins.}

\FloatBarrier

\newpage

\subsection{Fitted Signal and Background Yields}

Table~\ref{tab:rawresults} gives the fitted number of signal events
$\nsig$ in each individual $q^{2}$ and $\mx$ bin, along with
the fitted number of random combinatorial $\BB$ background events
$\nbb$ present in the signal enhanced region with $\mes > 5.27 \gevcc$.
The quoted uncertainties are statistical only.

\begin{table}[!h]
\caption{Fitted number of signal events $\nsig$ and
random combinatorial $\BB$ background events $\nbb$
present in the signal enhanced region with $\mes > 5.27 \gevcc$
by $q^{2} (\gevcccc)$ and $\mx (\gevcc)$ bin.}
\begin{center}
\centering\normalsize
\setlength{\tabcolsep}{8pt}
\begin{tabular}{cccccc}
\hline \hline
\noalign{\vskip 1mm}
&  & \multicolumn{2}{c}{\BToXee} & \multicolumn{2}{c}{\BToXmm} \\
Bin         & Range                 & \nsig                        & \nbb             & \nsig                             & \nbb             \\
\noalign{\vskip 1mm}
\hline
\noalign{\vskip 1mm}
$q^{2}_{0}$ & $1.0 < q^{2} < 6.0$   & $58.5^{+14.4}_{-13.5}$       & $348.8 \pm 22.2$ & $\: \: \: \: 8.6^{+11.2}_{-10.4}$ & $521.8 \pm 29.5$ \\
$q^{2}_{1}$ & $0.1 < q^{2} < 2.0$   & $60.4^{+11.9}_{-11.1}$       & $\: 95.3  \pm 12.4$ & $13.3^{+ 7.7}_{- 6.9}$            & $28.4  \pm 7.6 $ \\
$q^{2}_{2}$ & $2.0 < q^{2} < 4.3$   & $\! \! 20.8^{+ 9.4}_{- 8.5}$ & $168.2 \pm 15.3$ & $\! \! -1.9^{+ 6.9}_{- 5.8}$      & $50.5  \pm 8.6 $ \\
$q^{2}_{3}$ & $4.3 < q^{2} < 6.8$   & $25.4^{+10.3}_{- 9.5}$       & $181.3 \pm 16.2$ & $\: \: 6.0^{+ 9.0}_{- 8.3}$       & $\: 58.5  \pm 10.2$ \\
$q^{2}_{4}$ & $10.1< q^{2} <12.9$   & $59.1^{+14.8}_{-14.0}$       & $201.0 \pm 20.3$ & $\: 25.5^{+10.4}_{- 9.6}$         & $107.6 \pm 13.6$ \\
$q^{2}_{5}$ & $14.2< q^{2}$         & $\! \! 41.0^{+ 8.3}_{- 7.7}$ & $\: 40.2  \pm 10.0$ & $23.3^{+ 7.3}_{- 6.7}$            & $20.0  \pm 8.5 $ \\
\noalign{\vskip 1mm}
\hline
\noalign{\vskip 1mm}
$\mxone$    & $0.4 < \mx < 0.6 $ & $\! \! 63.0^{+ 9.9}_{- 9.2}$   & $\: \: 3.0   \pm 2.9 $ & $34.0^{+ 6.8}_{- 6.1}$             & $\: 2.0   \pm 1.9 $ \\
$\mxtwo$    & $0.6 < \mx < 1.0 $ & $68.1^{+11.5}_{-10.9}$         & $\: 38.0  \pm 8.9 $ & $22.0^{+ 7.3}_{- 6.7}$             & $12.1  \pm 6.2 $ \\
$\mxthree$  & $1.0 < \mx < 1.4 $ & $38.1^{+11.9}_{-11.3}$         & $168.1 \pm 17.9$ & $ \:  \: 7.3^{+ 9.0}_{- 8.4}$      & $\: 80.9  \pm 12.3$ \\
$\mxfour$   & $1.4 < \mx < 1.8 $ & $28.5^{+12.9}_{-12.3}$         & $483.9 \pm 28.6$ & $ \:  \:  \: 1.3^{+10.3}_{- 9.5}$  & $171.2 \pm 17.5$ \\
\noalign{\vskip 1mm}
\hline \hline
\end{tabular}
\label{tab:rawresults}
\end{center}
\end{table}

\FloatBarrier

\newpage

\subsection{Systematics Tables}

Tables~\ref{tab:addsystrates-ee}-\ref{tab:modelsystrates-mm} detail
the individual contributions to the branching fraction systematics.
Uncertainties quoted without a preceding ``$+$'' or ``$-$'' are $\pm$
symmetric.

\begin{table}[!h]
\centering
\caption{\BToXee\ branching fraction ``multiplicative'' systematic uncertainties.}
\label{tab:addsystrates-ee}
\vspace{0.25cm}
\setlength{\extrarowheight}{1.8pt}
\begin{tabular}{l|rrrrrr|rrrr}
\hline \hline
Variation                          & $q^{2}_{0}$   & $q^{2}_{1}$   & $q^{2}_{2}$   & $q^{2}_{3}$   & $q^{2}_{4}$   & $q^{2}_{5}$   & $\mxone$   & $\mxtwo$   & $\mxthree$   & $\mxfour$ \\
\hline
$N_{\BB}$                          & 0.012 & 0.018 & 0.004 & 0.004 & 0.007 & 0.003 & 0.004 & 0.007 & 0.010 & 0.011 \\
Tracking efficiency                & 0.031 & 0.049 & 0.011 & 0.011 & 0.018 & 0.009 & 0.011 & 0.019 & 0.026 & 0.030 \\
Particle Identification efficiency & 0.033 & 0.052 & 0.012 & 0.012 & 0.019 & 0.010 & 0.012 & 0.020 & 0.027 & 0.032 \\
$\KS$ efficiency                   & 0.004 & 0.006 & 0.001 & 0.001 & 0.001 & 0.001 & 0.001 & 0.002 & 0.002 & 0.002 \\
$\piz$ efficiency                  & 0.012 & 0.024 & 0.004 & 0.004 & 0.006 & 0.002 & 0.000 & 0.008 & 0.014 & 0.019 \\
BDT efficiency correction          & 0.004 & 0.006 & 0.001 & 0.001 & 0.009 & 0.004 & 0.001 & 0.002 & 0.003 & 0.004 \\
Monte Carlo statistics             & 0.002 & 0.009 & 0.002 & 0.002 & 0.003 & 0.001 & 0.001 & 0.001 & 0.002 & 0.006 \\
\hline
Total                              & 0.048 & 0.079 & 0.017 & 0.017 & 0.030 & 0.014 & 0.017 & 0.030 & 0.041 & 0.050 \\
\hline \hline
\end{tabular}
\end{table}

\begin{table}[!h]
\centering
\caption{\BToXmm\ branching fraction ``multiplicative'' systematic uncertainties.}
\label{tab:addsystrates-mm}
\vspace{0.25cm}
\setlength{\extrarowheight}{1.8pt}
\begin{tabular}{l|rrrrrr|rrrr}
\hline \hline
Variation                          & $q^{2}_{0}$   & $q^{2}_{1}$   & $q^{2}_{2}$   & $q^{2}_{3}$   & $q^{2}_{4}$   & $q^{2}_{5}$   & $\mxone$   & $\mxtwo$   & $\mxthree$   & $\mxfour$ \\
\hline
$N_{\BB}$                          & 0.004 & 0.011 & 0.001 & 0.002 & 0.005 & 0.004 & 0.004 & 0.005 & 0.004 & 0.001 \\
Tracking efficiency                & 0.009 & 0.027 & 0.002 & 0.005 & 0.011 & 0.007 & 0.008 & 0.011 & 0.010 & 0.003 \\
Particle Identification efficiency & 0.015 & 0.042 & 0.003 & 0.008 & 0.020 & 0.014 & 0.017 & 0.017 & 0.015 & 0.004 \\
$\KS$ efficiency                   & 0.001 & 0.004 & 0.000 & 0.001 & 0.002 & 0.001 & 0.001 & 0.002 & 0.001 & 0.000 \\
$\piz$ efficiency                  & 0.004 & 0.013 & 0.001 & 0.002 & 0.004 & 0.002 & 0.000 & 0.005 & 0.005 & 0.002 \\
BDT efficiency correction          & 0.002 & 0.005 & 0.000 & 0.001 & 0.010 & 0.007 & 0.002 & 0.002 & 0.002 & 0.001 \\
Monte Carlo statistics             & 0.001 & 0.005 & 0.000 & 0.001 & 0.003 & 0.001 & 0.001 & 0.001 & 0.001 & 0.001 \\
\hline
Total                              & 0.019 & 0.054 & 0.004 & 0.010 & 0.026 & 0.018 & 0.020 & 0.022 & 0.019 & 0.006 \\
\hline \hline
\end{tabular}
\end{table}

\begin{table}[!h]
\centering
\caption{\BToXee\ branching fraction ``additive'' systematic uncertainties.}
\label{tab:systrates-ee}
\vspace{0.25cm}
\setlength{\extrarowheight}{1.8pt}
\begin{tabular}{l|rrrrrr|rrrr}
\hline \hline
Variation                           & $q^{2}_{0}$   & $q^{2}_{1}$   & $q^{2}_{2}$   & $q^{2}_{3}$   & $q^{2}_{4}$   & $q^{2}_{5}$   & $\mxone$   & $\mxtwo$   & $\mxthree$   & $\mxfour$ \\
\hline
Signal $\mes$ pdf shape             & ${}^{+0.039}_{-0.019}$ & ${}^{+0.092}_{-0.092}$ & ${}^{+0.007}_{-0.014}$ & ${}^{+0.014}_{-0.008}$ & ${}^{+0.034}_{-0.034}$ & ${}^{+0.011}_{-0.005}$ & ${}^{+0.021}_{-0.007}$ & ${}^{+0.036}_{-0.012}$ & ${}^{+0.032}_{-0.017}$ & ${}^{+0.057}_{-0.057}$ \\
Signal $\lhr$ pdf shape             & ${}^{+0.077}_{-0.026}$ & ${}^{+0.159}_{-0.000}$ & ${}^{+0.055}_{-0.012}$ & ${}^{+0.006}_{-0.000}$ & ${}^{+0.142}_{-0.000}$ & ${}^{+0.002}_{-0.001}$ & ${}^{+0.002}_{-0.001}$ & ${}^{+0.050}_{-0.000}$ & ${}^{+0.127}_{-0.000}$ & ${}^{+0.474}_{-0.000}$ \\
Crossfeed pdf shape                 & ${}^{+0.032}_{-0.014}$ & ${}^{+0.074}_{-0.090}$ & ${}^{+0.026}_{-0.009}$ & ${}^{+0.015}_{-0.003}$ & ${}^{+0.067}_{-0.016}$ & ${}^{+0.000}_{-0.015}$ & ${}^{+0.002}_{-0.001}$ & ${}^{+0.013}_{-0.000}$ & ${}^{+0.055}_{-0.017}$ & ${}^{+0.201}_{-0.000}$ \\
Crossfeed normalization             & ${}^{+0.034}_{-0.026}$ & ${}^{+0.039}_{-0.022}$ & ${}^{+0.014}_{-0.014}$ & ${}^{+0.011}_{-0.009}$ & ${}^{+0.021}_{-0.019}$ & ${}^{+0.008}_{-0.008}$ & ${}^{+0.002}_{-0.001}$ & ${}^{+0.014}_{-0.013}$ & ${}^{+0.027}_{-0.033}$ & ${}^{+0.011}_{-0.025}$ \\
$\BB$ pdf shape                     & ${}^{+0.159}_{-0.123}$ & ${}^{+0.139}_{-0.096}$ & ${}^{+0.080}_{-0.052}$ & ${}^{+0.086}_{-0.070}$ & ${}^{+0.010}_{-0.038}$ & ${}^{+0.014}_{-0.000}$ & ${}^{+0.006}_{-0.005}$ & ${}^{+0.020}_{-0.019}$ & ${}^{+0.126}_{-0.099}$ & ${}^{+0.367}_{-0.351}$ \\
$udsc$ pdf shape                    & ${}^{+0.063}_{-0.050}$ & ${}^{+0.106}_{-0.094}$ & ${}^{+0.026}_{-0.021}$ & ${}^{+0.032}_{-0.034}$ & ${}^{+0.035}_{-0.026}$ & ${}^{+0.014}_{-0.012}$ & ${}^{+0.005}_{-0.006}$ & ${}^{+0.021}_{-0.025}$ & ${}^{+0.087}_{-0.089}$ & ${}^{+0.181}_{-0.231}$ \\
$udsc$ normalization                & ${}^{+0.019}_{-0.007}$ & ${}^{+0.059}_{-0.039}$ & ${}^{+0.009}_{-0.005}$ & ${}^{+0.011}_{-0.006}$ & ${}^{+0.019}_{-0.009}$ & ${}^{+0.008}_{-0.003}$ & ${}^{+0.003}_{-0.003}$ & ${}^{+0.010}_{-0.013}$ & ${}^{+0.030}_{-0.031}$ & ${}^{+0.055}_{-0.057}$ \\
Charmonium pdf shape                & ${}^{+0.055}_{-0.029}$ & ${}^{+0.020}_{-0.006}$ & ${}^{+0.019}_{-0.012}$ & ${}^{+0.032}_{-0.046}$ & ${}^{+0.115}_{-0.038}$ & ${}^{+0.005}_{-0.000}$ & ${}^{+0.006}_{-0.000}$ & ${}^{+0.034}_{-0.018}$ & ${}^{+0.126}_{-0.061}$ & ${}^{+0.231}_{-0.139}$ \\
Charmonium normalization            & ${}^{+0.041}_{-0.034}$ & ${}^{+0.022}_{-0.006}$ & ${}^{+0.024}_{-0.021}$ & ${}^{+0.041}_{-0.037}$ & ${}^{+0.075}_{-0.067}$ & ${}^{+0.012}_{-0.007}$ & ${}^{+0.025}_{-0.024}$ & ${}^{+0.046}_{-0.048}$ & ${}^{+0.093}_{-0.097}$ & ${}^{+0.107}_{-0.113}$ \\
\hline
Total                               & ${}^{+0.210}_{-0.147}$ & ${}^{+0.275}_{-0.192}$ & ${}^{+0.110}_{-0.066}$ & ${}^{+0.109}_{-0.099}$ & ${}^{+0.216}_{-0.100}$ & ${}^{+0.029}_{-0.023}$ & ${}^{+0.034}_{-0.026}$ & ${}^{+0.092}_{-0.064}$ & ${}^{+0.265}_{-0.183}$ & ${}^{+0.709}_{-0.465}$ \\
\hline \hline
\end{tabular}
\end{table}

\begin{table}[!h]
\centering
\caption{\BToXmm\ branching fraction ``additive'' systematic uncertainties.}
\label{tab:systrates-mm}
\vspace{0.25cm}
\setlength{\extrarowheight}{1.8pt}
\begin{tabular}{l|rrrrrr|rrrr}
\hline \hline
Variation                           & $q^{2}_{0}$   & $q^{2}_{1}$   & $q^{2}_{2}$   & $q^{2}_{3}$   & $q^{2}_{4}$   & $q^{2}_{5}$   & $\mxone$   & $\mxtwo$   & $\mxthree$   & $\mxfour$ \\
\hline
Signal $\mes$ pdf shape             & ${}^{+0.012}_{-0.007}$ & ${}^{+0.055}_{-0.018}$ & ${}^{+0.004}_{-0.004}$ & ${}^{+0.010}_{-0.006}$ & ${}^{+0.008}_{-0.008}$ & ${}^{+0.012}_{-0.012}$ & ${}^{+0.022}_{-0.022}$ & ${}^{+0.008}_{-0.015}$ & ${}^{+0.019}_{-0.019}$ & ${}^{+0.006}_{-0.004}$ \\
Signal $\lhr$ pdf shape             & ${}^{+0.030}_{-0.048}$ & ${}^{+0.000}_{-0.069}$ & ${}^{+0.029}_{-0.038}$ & ${}^{+0.000}_{-0.025}$ & ${}^{+0.024}_{-0.001}$ & ${}^{+0.003}_{-0.002}$ & ${}^{+0.000}_{-0.003}$ & ${}^{+0.000}_{-0.008}$ & ${}^{+0.000}_{-0.098}$ & ${}^{+0.378}_{-0.000}$ \\
Crossfeed pdf shape                 & ${}^{+0.018}_{-0.020}$ & ${}^{+0.077}_{-0.123}$ & ${}^{+0.029}_{-0.011}$ & ${}^{+0.010}_{-0.006}$ & ${}^{+0.039}_{-0.035}$ & ${}^{+0.018}_{-0.015}$ & ${}^{+0.003}_{-0.003}$ & ${}^{+0.033}_{-0.006}$ & ${}^{+0.077}_{-0.026}$ & ${}^{+0.136}_{-0.118}$ \\
Crossfeed normalization             & ${}^{+0.025}_{-0.020}$ & ${}^{+0.030}_{-0.008}$ & ${}^{+0.005}_{-0.002}$ & ${}^{+0.016}_{-0.013}$ & ${}^{+0.020}_{-0.017}$ & ${}^{+0.007}_{-0.006}$ & ${}^{+0.002}_{-0.001}$ & ${}^{+0.007}_{-0.007}$ & ${}^{+0.017}_{-0.039}$ & ${}^{+0.017}_{-0.011}$ \\
$\BB$ pdf shape                     & ${}^{+0.226}_{-0.121}$ & ${}^{+0.141}_{-0.063}$ & ${}^{+0.094}_{-0.045}$ & ${}^{+0.097}_{-0.054}$ & ${}^{+0.058}_{-0.009}$ & ${}^{+0.011}_{-0.000}$ & ${}^{+0.002}_{-0.005}$ & ${}^{+0.012}_{-0.013}$ & ${}^{+0.182}_{-0.104}$ & ${}^{+0.340}_{-0.199}$ \\
$udsc$ pdf shape                    & ${}^{+0.050}_{-0.050}$ & ${}^{+0.083}_{-0.065}$ & ${}^{+0.013}_{-0.007}$ & ${}^{+0.065}_{-0.078}$ & ${}^{+0.024}_{-0.021}$ & ${}^{+0.013}_{-0.012}$ & ${}^{+0.001}_{-0.000}$ & ${}^{+0.033}_{-0.029}$ & ${}^{+0.064}_{-0.070}$ & ${}^{+0.117}_{-0.136}$ \\
$udsc$ normalization                & ${}^{+0.030}_{-0.016}$ & ${}^{+0.059}_{-0.042}$ & ${}^{+0.207}_{-0.000}$ & ${}^{+0.022}_{-0.016}$ & ${}^{+0.021}_{-0.015}$ & ${}^{+0.003}_{-0.000}$ & ${}^{+0.005}_{-0.004}$ & ${}^{+0.012}_{-0.013}$ & ${}^{+0.032}_{-0.040}$ & ${}^{+0.069}_{-0.065}$ \\
Charmonium pdf shape                & ${}^{+0.069}_{-0.121}$ & ${}^{+0.030}_{-0.030}$ & ${}^{+0.032}_{-0.056}$ & ${}^{+0.056}_{-0.033}$ & ${}^{+0.019}_{-0.022}$ & ${}^{+0.003}_{-0.000}$ & ${}^{+0.002}_{-0.001}$ & ${}^{+0.024}_{-0.007}$ & ${}^{+0.026}_{-0.066}$ & ${}^{+0.306}_{-0.283}$ \\
Charmonium normalization            & ${}^{+0.134}_{-0.117}$ & ${}^{+0.083}_{-0.073}$ & ${}^{+0.085}_{-0.081}$ & ${}^{+0.102}_{-0.095}$ & ${}^{+0.039}_{-0.032}$ & ${}^{+0.004}_{-0.000}$ & ${}^{+0.007}_{-0.007}$ & ${}^{+0.032}_{-0.033}$ & ${}^{+0.098}_{-0.106}$ & ${}^{+0.246}_{-0.231}$ \\
Hadronic misidenti- \\
\hspace{2mm} fication pdf shape     & ${}^{+0.098}_{-0.087}$ & ${}^{+0.099}_{-0.085}$ & ${}^{+0.061}_{-0.063}$ & ${}^{+0.060}_{-0.054}$ & ${}^{+0.051}_{-0.044}$ & ${}^{+0.030}_{-0.026}$ & ${}^{+0.029}_{-0.028}$ & ${}^{+0.034}_{-0.035}$ & ${}^{+0.099}_{-0.109}$ & ${}^{+0.195}_{-0.187}$ \\
Hadronic misidenti- \\
\hspace{2mm} fication normalization & ${}^{+0.030}_{-0.027}$ & ${}^{+0.166}_{-0.111}$ & ${}^{+0.049}_{-0.038}$ & ${}^{+0.059}_{-0.021}$ & ${}^{+0.012}_{-0.021}$ & ${}^{+0.007}_{-0.012}$ & ${}^{+0.012}_{-0.003}$ & ${}^{+0.021}_{-0.022}$ & ${}^{+0.084}_{-0.071}$ & ${}^{+0.017}_{-0.078}$ \\
\hline
Total                               & ${}^{+0.299}_{-0.239}$ & ${}^{+0.292}_{-0.237}$ & ${}^{+0.260}_{-0.137}$ & ${}^{+0.187}_{-0.154}$ & ${}^{+0.107}_{-0.079}$ & ${}^{+0.043}_{-0.037}$ & ${}^{+0.040}_{-0.037}$ & ${}^{+0.076}_{-0.066}$ & ${}^{+0.268}_{-0.249}$ & ${}^{+0.699}_{-0.501}$ \\
\hline \hline
\end{tabular}
\end{table}



\begin{table}[!h]
\centering
\caption{\BToXee\ branching fraction model-dependent extrapolation systematic uncertainties.}
\label{tab:modelsystrates-ee}
\vspace{0.25cm}
\setlength{\extrarowheight}{1.8pt}
\begin{tabular}{l|rrrrrr|rr}
\hline \hline
Variation                         & $q^{2}_{0}$   & $q^{2}_{1}$   & $q^{2}_{2}$   & $q^{2}_{3}$   & $q^{2}_{4}$   & $q^{2}_{5}$   & $\mxthree$   & $\mxfour$ \\
\hline
Jetset tunings                    & ${}^{+0.060}_{-0.059}$ & ${}^{+0.011}_{-0.013}$ & ${}^{+0.010}_{-0.012}$ & ${}^{+0.011}_{-0.014}$ & ${}^{+0.001}_{-0.002}$ & ${}^{+0.031}_{-0.036}$ & ${}^{+0.037}_{-0.036}$ & ${}^{+0.075 }_{-0.077}$ \\
$\pm 50\%$ $N_{\piz}>1$           & $    0.249$            & $    0.047$            & $    0.038$            & $    0.025$            & $    0.002$            & $    0.130$            & $    0.030$            & $    0.051$             \\
$\pm 50\%$ $K$ multiplicity       & $    0.046$            & $    0.008$            & $    0.006$            & $    0.002$            & $    0.000$            & $    0.022$            & $    0.000$            & $    0.006$             \\
$\pm 50\%$ $\pip$ multiplicity    & $    0.196$            & $    0.036$            & $    0.028$            & $    0.012$            & $    0.000$            & $    0.100$            & $    0.024$            & $    0.080$             \\
$\pm 1 \sigma$ $\kmaybell$ BFs    & ${}^{+0.115}_{-0.129}$ & ${}^{+0.024}_{-0.026}$ & ${}^{+0.021}_{-0.023}$ & ${}^{+0.018}_{-0.018}$ & ${}^{+0.002}_{-0.002}$ & ${}^{+0.067}_{-0.073}$ & ${}^{+0.004}_{-0.005}$ & ${}^{+0.000}_{-0.000}$  \\
\hline
Total                             & ${}^{+0.346}_{-0.351}$ & ${}^{+0.065}_{-0.066}$ & ${}^{+0.053}_{-0.054}$ & ${}^{+0.035}_{-0.036}$ & ${}^{+0.003}_{-0.003}$ & ${}^{+0.181}_{-0.184}$ & ${}^{+0.053}_{-0.053}$ & ${}^{+0.121}_{-0.123}$  \\
\hline \hline
\end{tabular}
\end{table}

\begin{table}[!h]
\centering
\caption{\BToXmm\ branching fraction model-dependent extrapolation systematic uncertainties.}
\label{tab:modelsystrates-mm}
\vspace{0.25cm}
\setlength{\extrarowheight}{1.8pt}
\begin{tabular}{l|rrrrrr|rr}
\hline \hline
Variation                         & $q^{2}_{0}$   & $q^{2}_{1}$   & $q^{2}_{2}$   & $q^{2}_{3}$   & $q^{2}_{4}$   & $q^{2}_{5}$   & $\mxthree$   & $\mxfour$ \\
\hline
Jetset tunings                    & ${}^{+0.035}_{-0.041}$ & ${}^{+0.002}_{-0.003}$ & ${}^{+0.005}_{-0.006}$ & ${}^{+0.009}_{-0.012}$ & ${}^{+0.001}_{-0.002}$ & ${}^{+0.025}_{-0.020}$ & ${}^{+0.015}_{-0.014}$ & ${}^{+0.007 }_{-0.008}$ \\
$\pm 50\%$ $N_{\piz}>1$           & $    0.154$            & $    0.011$            & $    0.020$            & $    0.021$            & $    0.002$            & $    0.047$            & $    0.012$            & $    0.005$             \\
$\pm 50\%$ $K$ multiplicity       & $    0.029$            & $    0.002$            & $    0.003$            & $    0.002$            & $    0.000$            & $    0.008$            & $    0.000$            & $    0.001$             \\
$\pm 50\%$ $\pip$ multiplicity    & $    0.122$            & $    0.008$            & $    0.015$            & $    0.010$            & $    0.000$            & $    0.036$            & $    0.010$            & $    0.008$             \\
$\pm 1 \sigma$ $\kmaybell$ BFs    & ${}^{+0.027}_{-0.030}$ & ${}^{+0.002}_{-0.002}$ & ${}^{+0.004}_{-0.005}$ & ${}^{+0.007}_{-0.007}$ & ${}^{+0.001}_{-0.001}$ & ${}^{+0.015}_{-0.019}$ & ${}^{+0.001}_{-0.001}$ & ${}^{+0.000}_{-0.000}$  \\
\hline
Total                             & ${}^{+0.203}_{-0.205}$ & ${}^{+0.014}_{-0.014}$ & ${}^{+0.026}_{-0.026}$ & ${}^{+0.026}_{-0.027}$ & ${}^{+0.003}_{-0.003}$ & ${}^{+0.066}_{-0.065}$ & ${}^{+0.021}_{-0.021}$ & ${}^{+0.012}_{-0.013}$  \\
\hline \hline
\end{tabular}
\end{table}

\FloatBarrier

\newpage


\subsection{Fit Projections}

The pages following show the \BToXee\ and \BToXmm\
branching fraction fit projections for each $q^{2}$ and $\mx$ bin.

\clearpage

\clearpage

\clearpage

\clearpage

\end{document}

%% file: authors_sep2013.tex
%
\author{J.~P.~Lees}
\author{V.~Poireau}
\author{V.~Tisserand}
\affiliation{Laboratoire d'Annecy-le-Vieux de Physique des Particules (LAPP), Universit\'e de Savoie, CNRS/IN2P3,  F-74941 Annecy-Le-Vieux, France}
\author{E.~Grauges}
\affiliation{Universitat de Barcelona, Facultat de Fisica, Departament ECM, E-08028 Barcelona, Spain }
\author{A.~Palano$^{ab}$ }
\affiliation{INFN Sezione di Bari$^{a}$; Dipartimento di Fisica, Universit\`a di Bari$^{b}$, I-70126 Bari, Italy }
\author{G.~Eigen}
\author{B.~Stugu}
\affiliation{University of Bergen, Institute of Physics, N-5007 Bergen, Norway }
\author{D.~N.~Brown}
\author{L.~T.~Kerth}
\author{Yu.~G.~Kolomensky}
\author{M.~J.~Lee}
\author{G.~Lynch}
\affiliation{Lawrence Berkeley National Laboratory and University of California, Berkeley, California 94720, USA }
% ...
\collaboration{The \babar\ Collaboration}
\noaffiliation

%% file: Xsll_arxiv_final_version.bbl
\begin{thebibliography}{99}
\bibitem{CC}
  The use of charge conjugate reactions is implied unless otherwise indicated.
\end{thebibliography}
